\documentclass[a4paper,11pt,twocolumn]{article}
\usepackage[switch]{lineno} 

\usepackage[utf8]{inputenc}
\usepackage{comment}

\usepackage[normalem]{ulem}
\usepackage{color}

\usepackage{soul}
\usepackage{amsmath,amssymb}
\usepackage{supertabular}
\usepackage{tabularx}

\usepackage{orcidlink}
\usepackage{comment}

\usepackage{overpic}   

\usepackage{url}

\RequirePackage[T1]{fontenc}

\RequirePackage{graphicx}
\RequirePackage{mathptmx}      
\RequirePackage{flushend}
\RequirePackage[numbers,sort&compress]{natbib}

\usepackage[blocks]{authblk} 
\usepackage{orcidlink}

\title{Beam test studies for a SiPM-based RICH detector prototype for the future ALICE~3 experiment}

\author[a]{A.~R.~Altamura \orcidlink{0000-0001-8048-5500}}
\author[a]{L.~Congedo \orcidlink{0000-0003-4536-4644}}
\author[a]{G.~De~Robertis \orcidlink{0000-0001-8261-6236}}
\author[a,d]{D.~Di~Bari}
\author[b]{A.~Di~Mauro \orcidlink{0000-0003-0348-092X}}
\author[a,d]{M. Giliberti \orcidlink{0009-0007-2835-2963}}
\author[c]{J.~O. Guerra-Pulido \orcidlink{0000-0002-1994-3573}}
\author[a]{F.~Licciulli \orcidlink{0000-0002-6955-0321}}
\author[a]{L.~Lorusso \orcidlink{0000-0002-2549-4401}}
\author[b]{P.~Martinengo \orcidlink{0000-0003-0288-202X}}
\author[a,1]{M.~N.~Mazziotta \orcidlink{0000-0001-9325-4672}}
\author[a]{E.~Nappi \orcidlink{0000-0003-2080-9010}}
\author[a,d,2]{N.~Nicassio \orcidlink{0000-0002-7839-2951}}
\author[c]{G.~Pai\'c \orcidlink{0000-0003-2513-2459}}
\author[a,d]{G.~Panzarini \orcidlink{0000-0002-2586-1021}}
\author[a]{R.~Pillera \orcidlink{0000-0003-3808-963X}}
\author[a,d]{G.~Volpe \orcidlink{0000-0002-2921-2475}}

\affil[a]{Istituto Nazionale di Fisica Nucleare (INFN), Sezione di Bari, via Orabona 4, I-70126 Bari, Italy}
\affil[b]{CERN, the European Organization for Nuclear Research, Esplanade des Particules 1, 1211 Geneva, Switzerland}
\affil[c]{Instituto de Ciencias Nucleares, Universidad Nacional Autónoma de México, Circuito Exterior s/n, Ciudad Universitaria, Col. UNAM, Coyoacán, 04510 Ciudad de México, México}
\affil[d]{Dipartimento di Fisica dell'Universit\`a e del Politecnico di Bari, via Amendola 173, I-70126 Bari, Italy}

\affil[1]{e-mail: mazziotta@ba.infn.it (corresponding author)}
\affil[2]{e-mail: nicola.nicassio@ba.infn.it (corresponding author)}
\date{}

\begin{document}

\maketitle

\begin{abstract}
The ALICE Collaboration is proposing a completely new apparatus, ALICE~3, for the LHC Runs~5 and beyond.
In this context, a key subsystem for high-energy charged particle identification will be a proximity-focusing ring-imaging Cherenkov detector using aerogel as radiator and silicon photomultipliers (SiPMs) as photon sensors.
We assembled a small-scale prototype instrumented with Hamamatsu S13352 and S13361-3075AE-08 SiPM arrays, readout by custom boards equipped with front-end Petiroc 2A ASICs.
The Cherenkov radiator consisted of a 2 cm thick hydrophobic aerogel tile with a refractive index of 1.03 separated from the SiPM plane by a 23 cm expansion gap. The prototype was successfully tested in a campaign at the CERN PS T10 beam line with the goal of validating the design bRICH specifications in
terms to achieve the target separation power. We measured a single photon angular resolution of 3.8~mrad at the Cherenkov angle saturation value of 242~mrad, as well as the expected scaling of the angular resolution with the increasing number of detected photons. We also studied the contribution of uncorrelated and correlated background sources with respect to the signal and proved the effectiveness of time matching between charged tracks and photon hits to achieve efficient suppression of the SiPM dark count rate background.  In this paper, the detector concept, the description of the tested prototype layout and the main beam test results are reported.
\end{abstract}

\section{Introduction}
\label{sec:intro}

The ALICE Collaboration is proposing a completely new apparatus, ALICE~3 \cite{ALICE3_loi}, for LHC Runs 5 and beyond with the aim of studying the properties of the quark-gluon plasma (QGP) formed in ultra-relativistic heavy-ion collisions beyond current limits \cite{ALICE_physics}. The ALICE~3 physics program includes precise measurements of dielectron emission and low-$p_{T}$ heavy-flavour production to probe the early evolution of QGP as well as its effects on the quark diffusion, thermalization and hadronization mechanisms. A key requirement for the ALICE~3 physics programme is extensive particle identification (PID) of $e^{\pm}$, $\mu^{\pm}$, $\pi^{\pm}$, $K^{\pm}$, $p$ and $\bar{p}$ with different dedicated subsystems providing acceptance over eight units of pseudorapidity ($|\eta|<4$). The design target of the ALICE~3 PID system is to ensure a better than 3$\sigma$ $e$/$\pi$, $\pi$/$K$ and $K$/$p$ separation for momenta up to 2 GeV/$c$, 10 GeV/$c$ and 16 GeV/$c$, respectively.
In order to match this goal, the baseline ALICE~3 PID system includes both ultra-fast time-of-flight (TOF) and ring-imaging Cherenkov (RICH) subsystems covering both the barrel ($|\eta|<2$) and the end-cap ($2<|\eta|<4$) regions. 

The required performance in the barrel RICH (bRICH) is achieved using a radiator with a refractive index $n~=~1.03$ and requiring an overall ring angular resolution better than 1.5~mrad at Cherenkov angle saturation.
The radiator refractive index is constrained by the need to ensure continuity  between the TOF PID upper momentum limits and the radiator momentum thresholds for Cherenkov emission, preventing the use of $n$ values below 1.03. Larger values of $n$ are ruled out due to the reduced separation power at larger momenta.
Dedicated simulation studies performed in the Geant4 framework \cite{AGOSTINELLI2003250} show that
the required resolution is expected to be achieved using a proximity-focusing layout \cite{Eugenio_RICH}, with 2 cm thick aerogel radiator tiles separated by an expansion gap larger than 20 cm from a photodetector surface segmented in $2\times2$~mm$^2$ cells and providing a single photon detection efficiency above 40\% at wavelengths close to 400 nm. 

The radiator is based on aerogel since it is the only viable option with the required optical properties~\cite{Eugenio_aerogel_RICH}. The use of aerogel for RICH applications is widespread in high-energy physics experiments~\cite{HERMES_RICH,LHCb_RICH,AMS_RICH,BELLE2_RICH,CLAS12_RICH}.
The use of aerogel is also envisaged for the RICH subsystems at future experiments, such as ePIC at the Electron Ion Collider~\cite{ePID_mRICH,ePID_dRICH}.

A projective geometry with modules featuring a radial thickness of about 30 cm placed at about 1 m radius from the beam axis and oriented toward the nominal collision vertex is considered for the bRICH. This results in an overall photodetector area of about 25 m$^2$. The detector is embedded in a superconducting magnet providing a solenoidal magnetic field of 2~T. For operation in central Pb-Pb collisions, a charged track multiplicity $\text{d}N_{\text{ch}}/\text{d}\eta>2000$ is expected \cite{ALICE_charged_track_multiplicity} , requiring for efficient pattern recognition in an environment with hundreds of tracks reaching the radiator per m$^{2}$.

As a further requirement, a single photon time resolution better than a few hundred picoseconds is needed to disentangle signal photon hits from background hits and photon hits from different tracks in the same event, especially in the high charged track multiplicity environment of central Pb-Pb collisions. By also accounting operation in a 2~T solenoidal magnetic field, the nowadays available photon sensor technology option is represented by SiPMs. 
Fluence levels of the order of 10$^{11}$ 1-MeV n$_{eq}$/cm$^2$ are expected to be 
integrated at the bRICH photodetector level throughout the full ALICE~3 operation according to state-of-the-art simulations.
Dedicated R\&D studies are therefore ongoing to prove the effectiveness of cooling the SiPMs down to -40~$^{\circ}$C by using bi-phase CO$_2$ microchannels \cite{CO2_cooling_paper} associated by annealing cycles~\cite{passive_annealing_paper,active_annealing_paper} to mitigate the increasing SiPM dark count rate (DCR) and validate the stability of the bRICH reconstruction performance up to such radiation levels.

A first prototype of the proposed RICH layout was already tested on beam in October 2022 with excellent results \cite{Nicassio_proceeding_IWASI}.
In this paper, we present the  results from the latest beam test at the CERN-PS T10 facility in October 2023, for a setup based on currently-off-the-shelf components with the goals of proving the achievable angular resolution, validating the bRICH geometry and studying the effects of timing on the SiPM background suppression.

\section{Test beam set-up and DAQ system}
\label{section_testbeam}

\begin{figure}
\centering
\includegraphics[width=0.95\columnwidth,height=0.22\textheight]{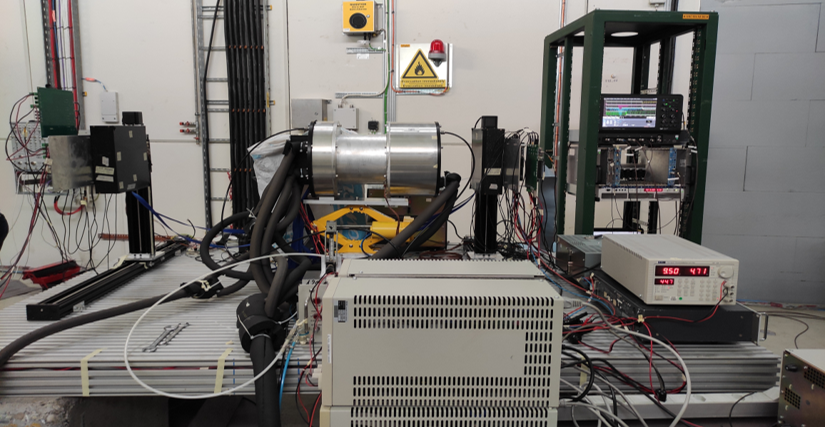}
\includegraphics[width=0.95\columnwidth, height=0.23\textheight]{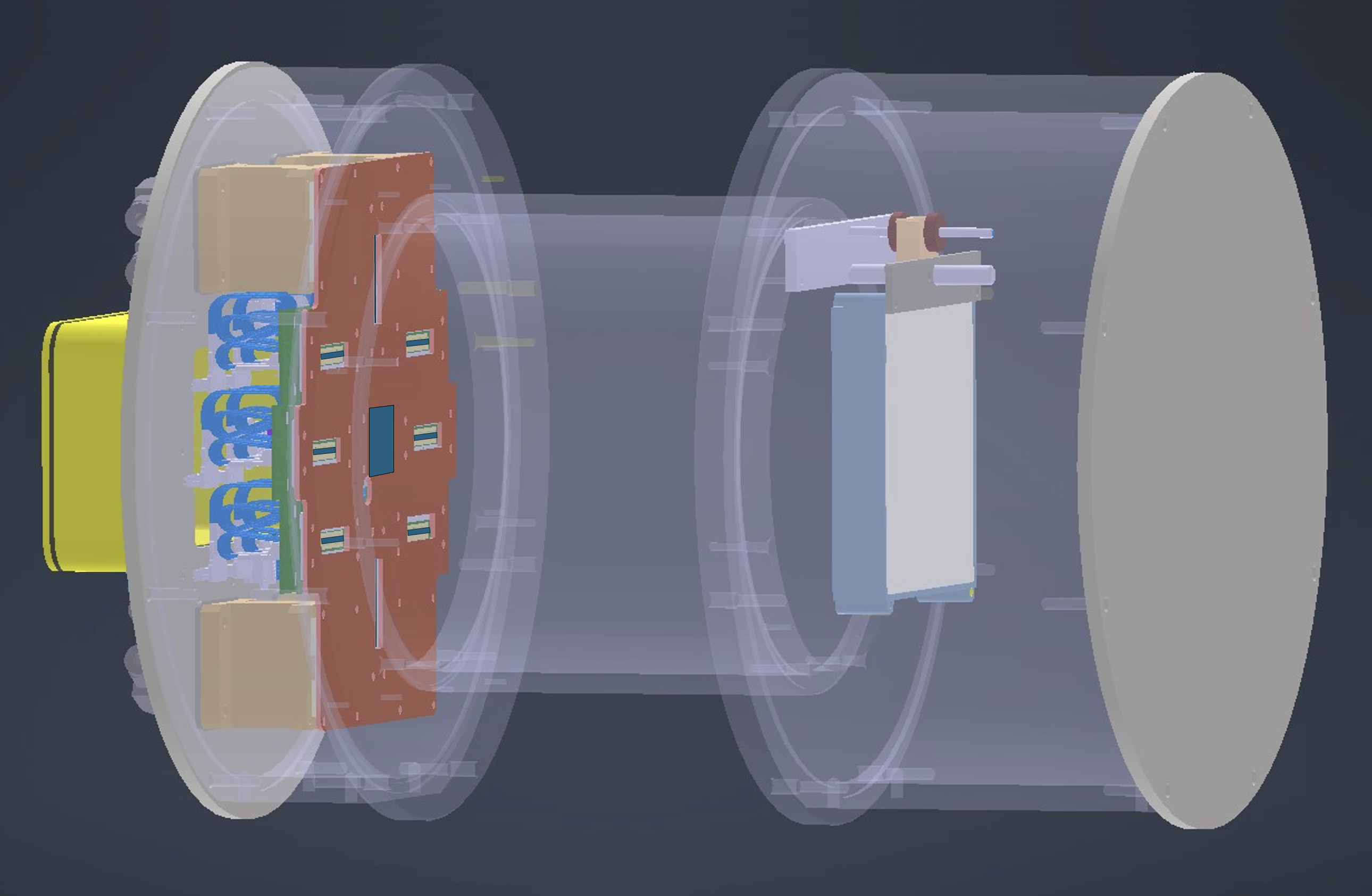}
\includegraphics[width=0.95\columnwidth,height=0.245\textheight]{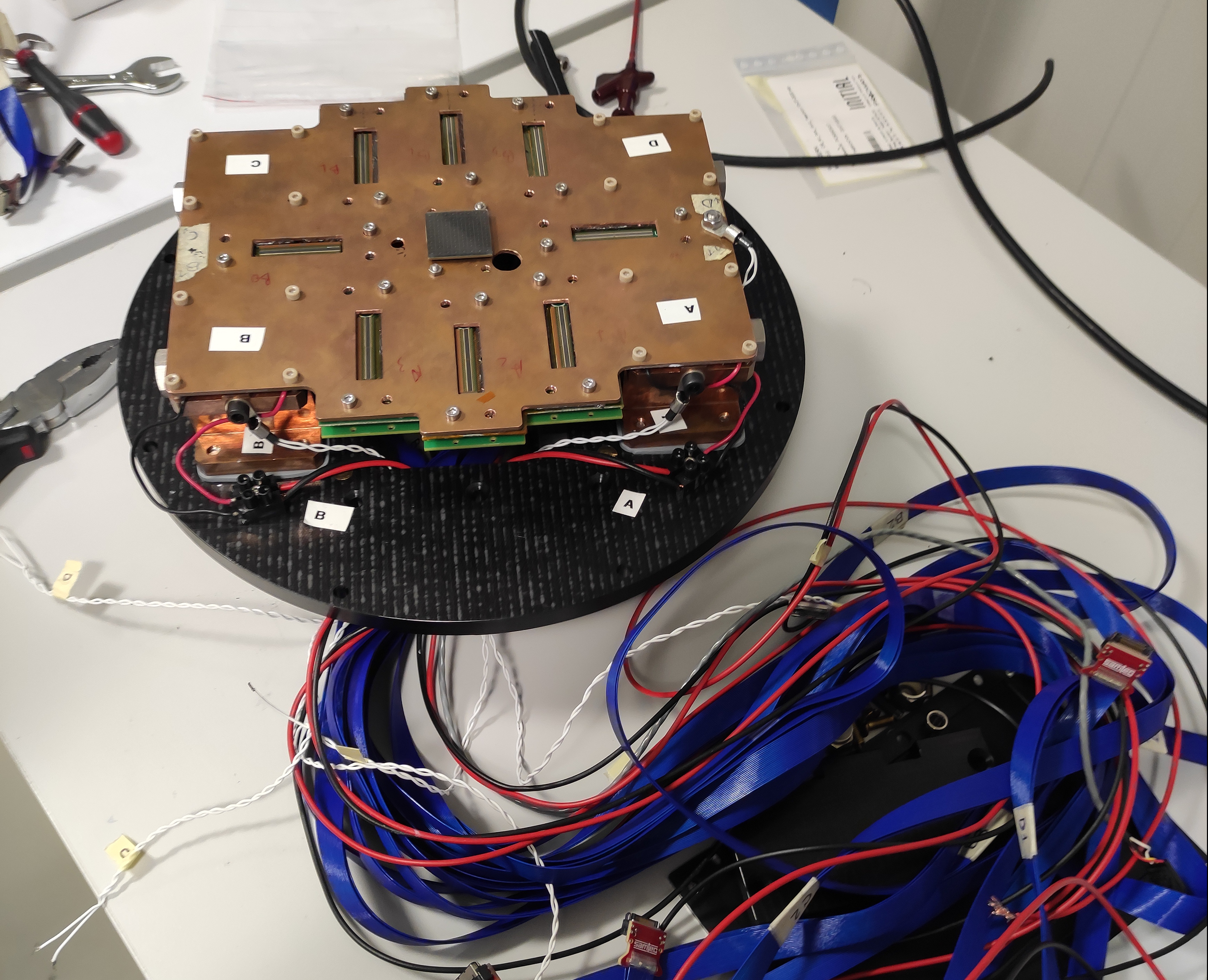}
\caption{Top panel: Photo of the beam test set-up at CERN PS T10 line on Oct, 2023. The beam enters from the right side. The black boxes upstream and downstream the aluminium vessel in the middle house  thin plastic scintillator tiles and a X-Y fiber tracker module~\cite{MAZZIOTTA2022167040}. Middle panel: CAD view of the RICH vessel with the aerogel tile and the endcap photon detector plate. Bottom panel: photon detector plate with S13552 SiPM linear array and downstream SiPM S13661-3075AE-08 array.}
\label{fig:st2023}
\end{figure}

A picture of the set-up is shown in Fig.~\ref{fig:st2023}. The RICH prototype is housed in a cylindrical vessel (middle panel of Fig.~\ref{fig:st2023}). The 11$\times$11$\times$2~cm$^3$ hydrophobic aerogel tile radiator was provided by Aerogel Factory Co. Ltd (Chiba, Japan)~\cite{ADACHI_Aerogel} with a refractive index n $=$ 1.03 and a measured transmission length of about 5.1 cm, resulting from a scattering length of about 6.7 cm and an absorption length of about 21.6 cm, at 400~nm wavelength.

The tile is separated by a 23 cm gap from a copper plate equipped with SiPM arrays mounted on custom printed circuit boards (PCBs). 
A central HPK S13361-3075AE-08 64-channel SiPM array made of SiPMs having a 3$\times$3~mm$^{2}$ active area and 3.2$\times$3.2~mm$^{2}$ pitch~\cite{b_SiPM_S13361_3050AE08_datasheet,b_SiPM_S13360_3075_datasheet} is used for charged particle detection and timing.
A 1~mm thick fused silica window, acting as a second Cherenkov radiator, is directly coupled to the array for achieving a more precise information on the event timing by measuring the detection time of the cluster of Cherenkov photons emitted by the fused silica window.
The Cherenkov photons produced in the areogel tile are detected with eight SiPM HPK S13552~128-channel linear arrays having a $0.23 \times 1.625$~mm$^{2}$ active area packaged with a 0.25~mm linear pitch ~\cite{b_SiPM_S13552_datasheet}.
The centers of each HPK S13552 are located at about 6.5 cm of radius with respect to the center of the HPK S13361-3075AE-08 array (bottom panel of Fig.~\ref{fig:st2023}).

The SiPMs are cooled down to -5~$^\circ$C to reduce the DCR. The cooling system is based on Peltier cells interfaced to low-temperature water heat pipes.  In particular, the copper plate on which SiPM arrays are assembled serves as a cooling plate. On the opposite side with respect to the SiPM arrays, a Peltier is installed at each of four corners of the copper plate in a sandwich configuration with a thermal paste between the SiPM cooling plate and a copper cooling-pipe block in which water at 5~$^\circ$C is circulating. The resulting cooling system enables to keep the SiPM temperature below 0~$^\circ$C.

The vessel is flushed with argon gas to prevent condensation by keeping the relative humidity below 2.5\%, corresponding to a dew point temperature inside the cylinder of approximately -25~$^\circ$C,  significantly lower than the SiPM operation temperature.

Analog temperature sensors TT4-10KC3-T125-M5-500~\cite{ntc10k} are assembled on the edges of the copper plate to monitor its temperature by using Raspberry Pi 3 with ADS1115 16-Bit ADCs \cite{ads1115}. In addition, humidity sensors SHT31-D~\cite{sht31} are used to monitor both the room and the vessel humidity.

\begin{figure}[!tb] 
\centering
\centerline{\includegraphics[width=0.68\columnwidth]{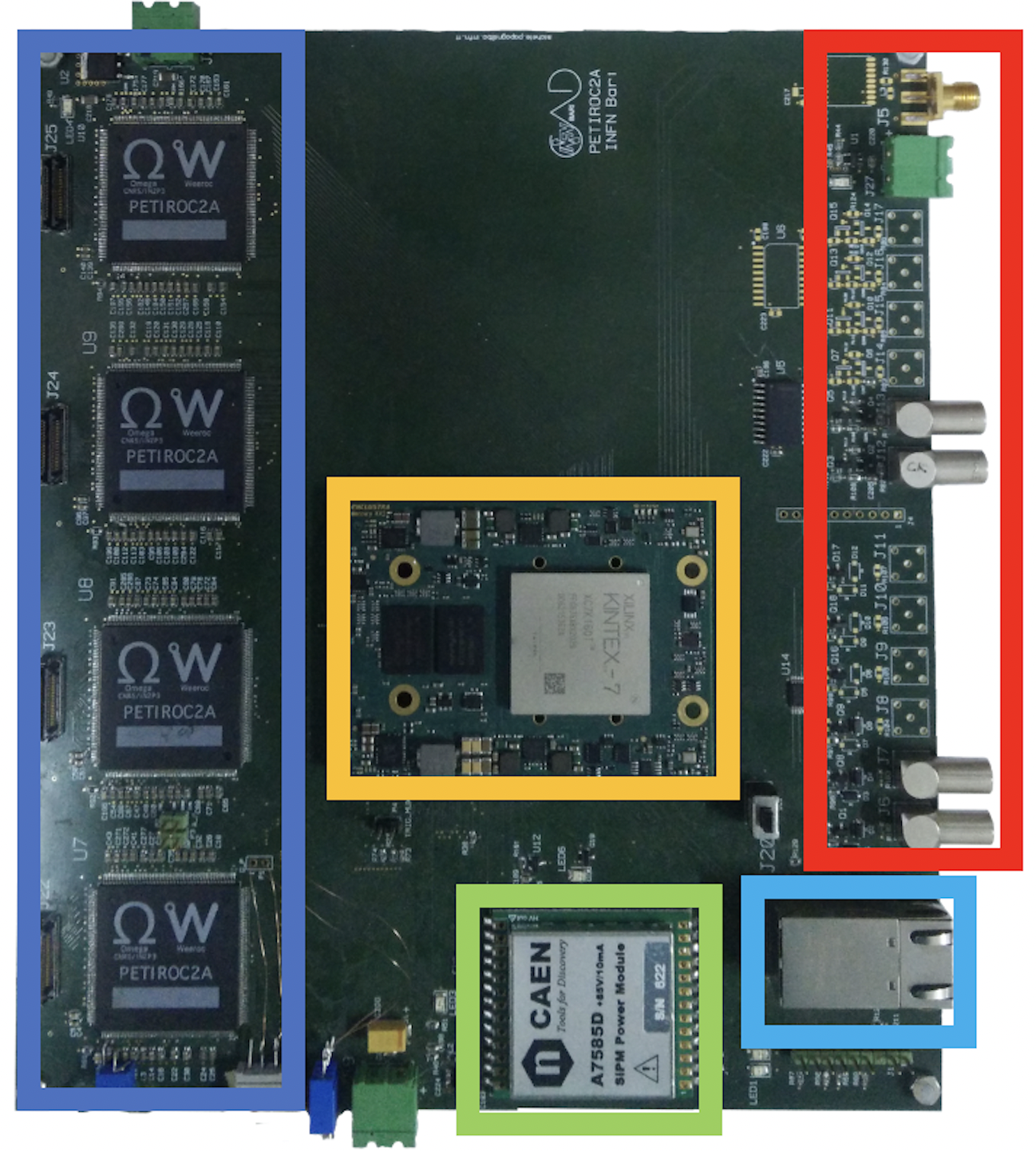}}
\vspace{1em}
\centerline{\includegraphics[width=0.78\columnwidth]{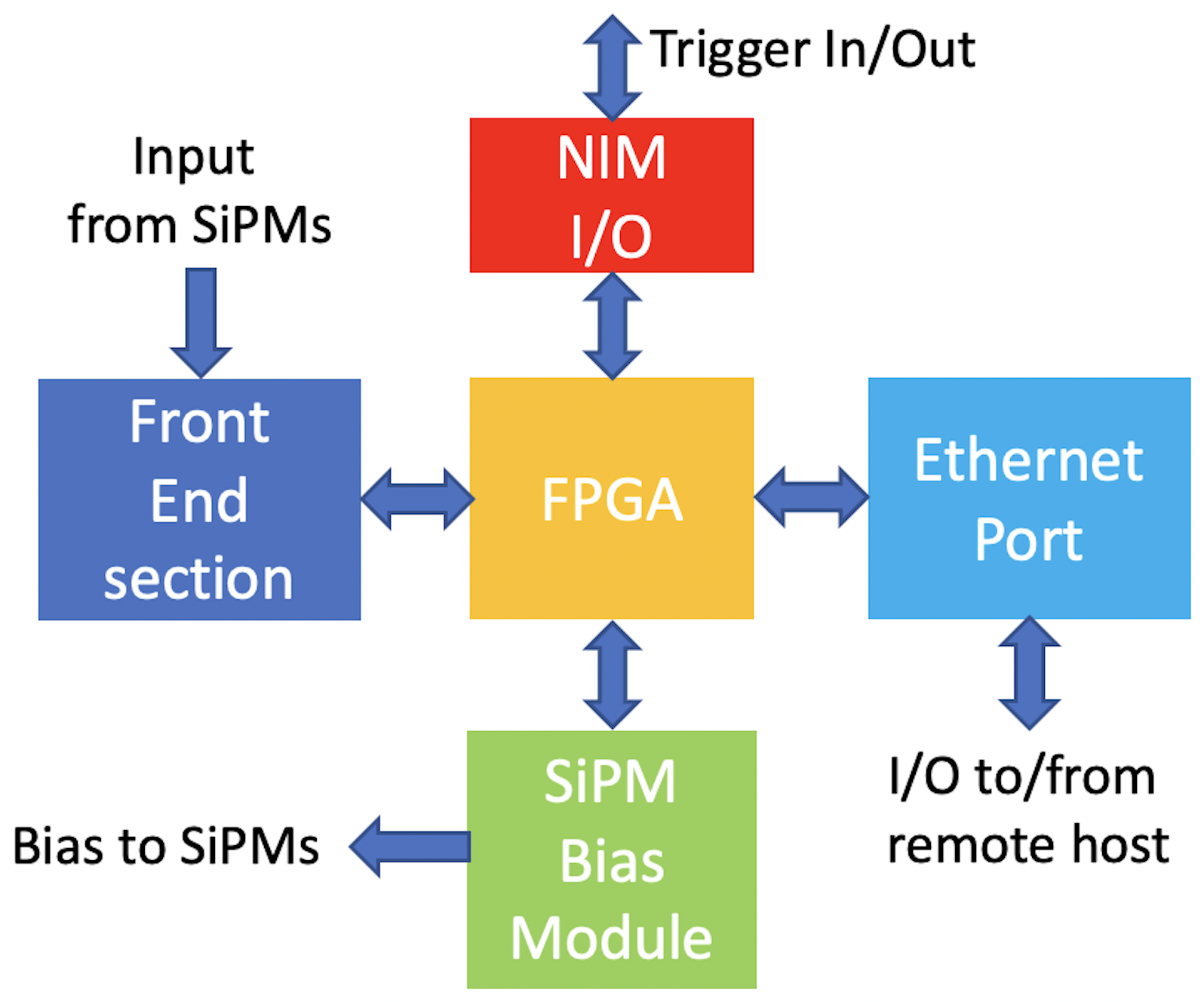}}
\caption{Picture (top) and block diagram (bottom)
of one of the boards used for data acquisition~\cite{MAZZIOTTA2022167040}.}
\label{fig:reading_interface}
\end{figure}

As for charged particle tracking and beam monitoring, the setup includes two X-Y tracker modules based on staggered round plastic scintillating fibers coupled with HPK S13552 128-channel SiPM arrays~\cite{MAZZIOTTA2022167040}.
The two tracker modules are mounted upstream and downstream the median plane of the aerogel tile respectively at 31~cm and 94~cm and are operated at room temperature. The distances are chosen to provide a precise information of the impact point of the particle onto the aerogel tile thus reducing the error in the reconstruction of aerogel Cherenkov photon emission angle.
The upstream module was assembled with fibers with 500 $\mu$m of diameter and readout on both sides, while the downstream one was assembled with 750 $\mu$m fibers and readout on one side. 

The resulting contribution of the tracking uncertainty on the Cherenkov angle resolution is expected to be smaller than 1~mrad.

All the S13552 arrays are bonded on one side of custom PCBs~\cite{MAZZIOTTA2022167040}, with the 128 channels of each array arranged into 32 groups, each consisting of 4 adjacent channels, resulting in an effective readout pitch of 1 mm~\cite{PILLERA2023167962}. 
Each of these groups of channels is routed to a Samtec LSHM-120 multi-channel connector~\cite{lshm} bonded on the opposite side of the PCBs. The S13361-3075AE-08 array is instead plugged with two Samtec ST4-40 connectors~\cite{st4} to a different custom PCB with two Samtec SS4-40 connectors~\cite{ss4} and all channels are routed to two Samtec LSHM-120 multi-channel connectors bonded on the opposite side.

The analog SiPM signals are routed from inside the vessel to the front-end electronics placed outside the vessel by means of 1 meter long high-speed 50 Ohm multi-channel Samtec HLCD-20 (40 channels) cables~\cite{hlcd}. The SiPM bias voltages are provided by means of 4 HLCD channels. All the SiPM PCBs are equipped with 1-wire digital temperature sensors DS18B20~\cite{ds18b20}, each routed by means of a HLCD channel~\cite{MAZZIOTTA2022167040}. 
The temperature of the copper plate and the ring array sensors was uniform within 1~$^\circ$C around -5~$^\circ$C, while the temperature on the central array sensor was stable around 0~$^\circ$C during data taking.

The SiPM readout interface is shown in Fig.~\ref{fig:reading_interface}. 
SiPM signals are readout by 32-channel Petiroc 2A~\cite{Ahmad:2018lxl} front-end ASICs, featuring a 10-bit ADC for charge measurements and a 37 ps-bin TDC for timing measurements, controlled by custom data acquisition boards developed by INFN Bari~\cite{MAZZIOTTA2022167040,PILLERA2023167962, 10164306}.
Each board hosts up to four Petiroc 2A ASICs, one CAEN A7585D SiPM voltage module~\cite{a7585} and one Kintex~-~7 FPGA mounted on a Mercury+KX2 module~\cite{b_Xilinx_FPGA} to configure the ASICs, the trigger configuration and its data acquisition.
Four different boards are used: one for the two ASICs used to readout the central array; two for the eight ASICs used to readout the ring arrays and one for the four ASICs used to readout the fiber tracker SiPM arrays.  

The boards are configured to operate in a master-slave configuration, with board reading the tracker modules as master unit providing a shared 40 MHz clock, trigger signal, and event tag to the slave units.
For each trigger event, a 550~ns time window for data acquisition is opened.
Within this time window, the system records the fired channels, along with their corresponding ADC and TDC readings for subsequent analysis. More details about the read-out and DAQ system can be found in
~\cite{MAZZIOTTA2022167040,PILLERA2023167962,10164306}.

The measurements were performed with both negative charged beam at 10~GeV/$c$ momentum and positive charged beam at 8~GeV/$c$ momentum. The negative charged beam  at 10~GeV/$c$ is mainly composed of pions, while the positive charged beam at 8~GeV/$c$ momentum is dominated by protons and pions.
For data taking, the system was aligned so that the 11 $\times$ 11 cm$^2$ surface of the aerogel tile and the plate equipped with SiPMs were normal to the beam. The CERN-PS T10 beam scintillator was used for external triggering.

\section{Charge and time calibration}
\label{sec:cal}

For each event we evaluated the actual ADC counts of individual channels by subtracting the corresponding pedestal values. We identified the peaks in the ADC charge distribution corresponding to photo-electrons (PEs), then we calculated the ADC-to-PE calibration constants. 

The hit time-of-arrival in the Petiroc 2A ASIC is done in two-step. A coarse-time ({\tt CT}) counter with a 40~MHz clock is used as reference and digitized with a 9-bits counter. A ramp-based Time-to-Amplitude converter (TAC) is used to interpolate the fine-time ({\tt FT}) between two coarse time edges and digitized with 10-bits counter. For each Petiroc 2A channel the discriminator output is used as start for the TAC ramp that is stopped by the following coarse counter signal clock running at 40 MHz. The time information $\text{t}(\text{ns})$ is reconstructed from the TDC values combining the 9-bits and 10-bits counters as follows:

\begin{equation}
\text{t}(\text{ns}) = \text{P}_{\text{40}} \times \left ( \texttt{CT} + 1 - \frac{\texttt{FT} - \texttt{FT}_{\text{min}}}{\texttt{FT}_{\text{max}}-\texttt{FT}_{\text{min}}} \right ) 
\end{equation}
where $\text{P}_{\text{40}}=25$ ns is the period of the
reference 40~MHz clock, $\texttt{FT}_{\text{min}}$ and $\texttt{FT}_{\text{max}}$ are the minimum and the maximum observed fine-time values for each channel, respectively~\footnote{The effective range of the 10-bits $\texttt{FT}$ counters typically starts at about 300 DAC units and ends at about 1000 DAC units. Consequently, the LSB resolution is slightly worse than the nominal 25~ps value due to the reduced dynamic range of the counters.}. More details about the calibration method are reported in ~\cite{MAZZIOTTA2022167040,PILLERA2023167962,10164306}.

Measured times are finally corrected for both the channel-by-channel offset, which is mainly due to signal routing and cabling through the ASICs, and the time-walk effect according to the observed number of PEs. The relative offsets among the signals in the arrays due to the nominal time-of-flight of the impinging charged particles and the emitted Cherenkov photons are also subtracted.

The over-voltage (OV) for SiPM operation at $25^\circ$C temperature was set to or slightly above the nominal 3.0$-$3.5~V excess voltage recommended for the HPK S13552 type sensors and 5~V for the HPK S13361 type sensors.
To increase the single Cherenkov photon detection efficiency when the cooling was on, we increased the average OV to about 6.3~V for the central array operated down to 0~$^\circ$C and 4.6~V for the ring linear arrays operated down to -5~$^\circ$C ~\footnote{To improve the charged particle timing performance, we operated the central array at a larger excess voltage with respect to the ring arrays.}.

The effective thresholds for data acquisition were optimized to efficiently detect the single PE signals from Cherenkov photons emitted in aerogel and the multi-PE signals from both Cherenkov photons emitted in the central array fused silica window and from scintillation photons in the tracker modules, respectively, while keeping the SiPM DCR noise level as low as possible.

\section{Test beam analysis and results}
\label{sec:analysis}

In the current analysis, we assumed a reference system with the positive Z-axis along the beam direction and the X and Y axes in the transverse plane following the right-hand rule. 
We selected events requiring that the SiPM channels with the maximum charge (cluster seeds) observed in the fiber tracker planes and in the central array had at least six PEs. In addition, we required that the position of those channels were within a central 8$\times$9~mm$^2$ fiducial area. We also required the offset-corrected times of those channels to be within a 8 ns window to suppress the background due to wrong or ghost hits in multi-particle events.

The charged particle Z coordinates were taken as the nominal position of the detectors along the beam line. The X-Y coordinates were assumed according to the cluster positions. The adjacent fired channels/pixel to the cluster seeds were associated to the clusters. Then the cluster positions were calculated averaging the nominal X-Y coordinates of the channels in the cluster each weighted with its corresponding number of PEs.
Finally, a three-point least-squares straight line fit was performed to extract the track parameters.

\begin{figure}[!t]
\centering
\includegraphics[width=0.95\columnwidth]{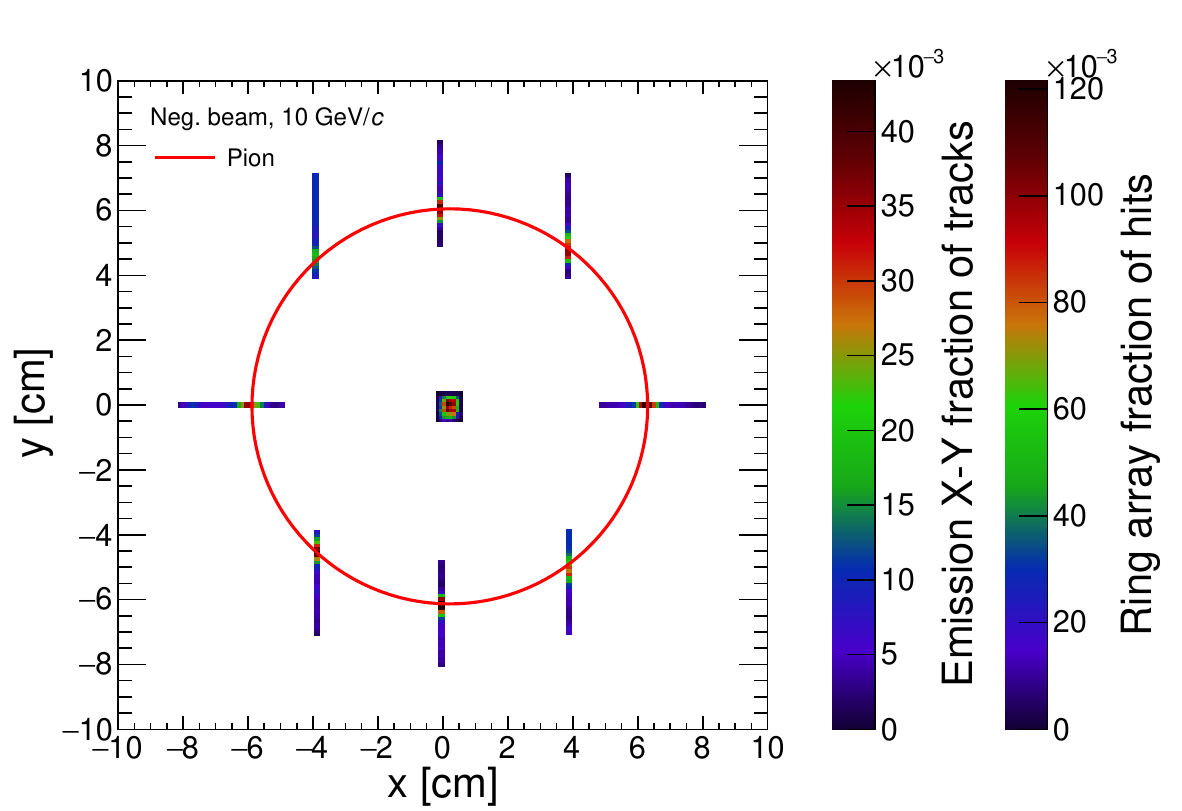}
\includegraphics[width=0.95\columnwidth]{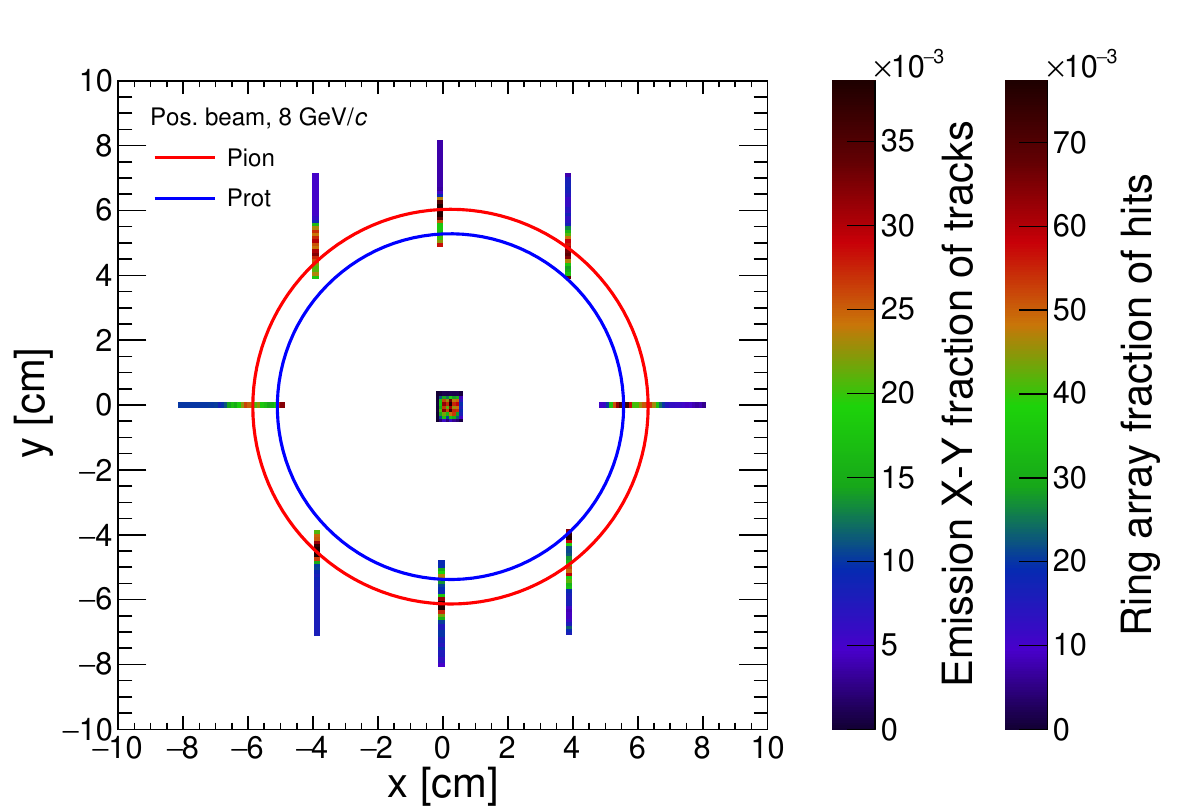}
\caption{
Spatial distribution of the extrapolated X-Y emission points in the median plane of the aerogel tile (central region) and of the hits in the eight ring arrays (ring region) for the negative charged beam at 10 GeV/$\it{c}$ momentum (top panel) and the positive charged beam at 8 GeV/$\it{c}$ momentum (bottom panel). For guiding the eyes, the expected Cherenkov rings from pions and protons centered in the most probable X-Y emission point are also shown.}
\label{fig:hit_maps}
\end{figure}

\begin{figure}[!h!t]
\centering
\includegraphics[width=0.95\columnwidth]{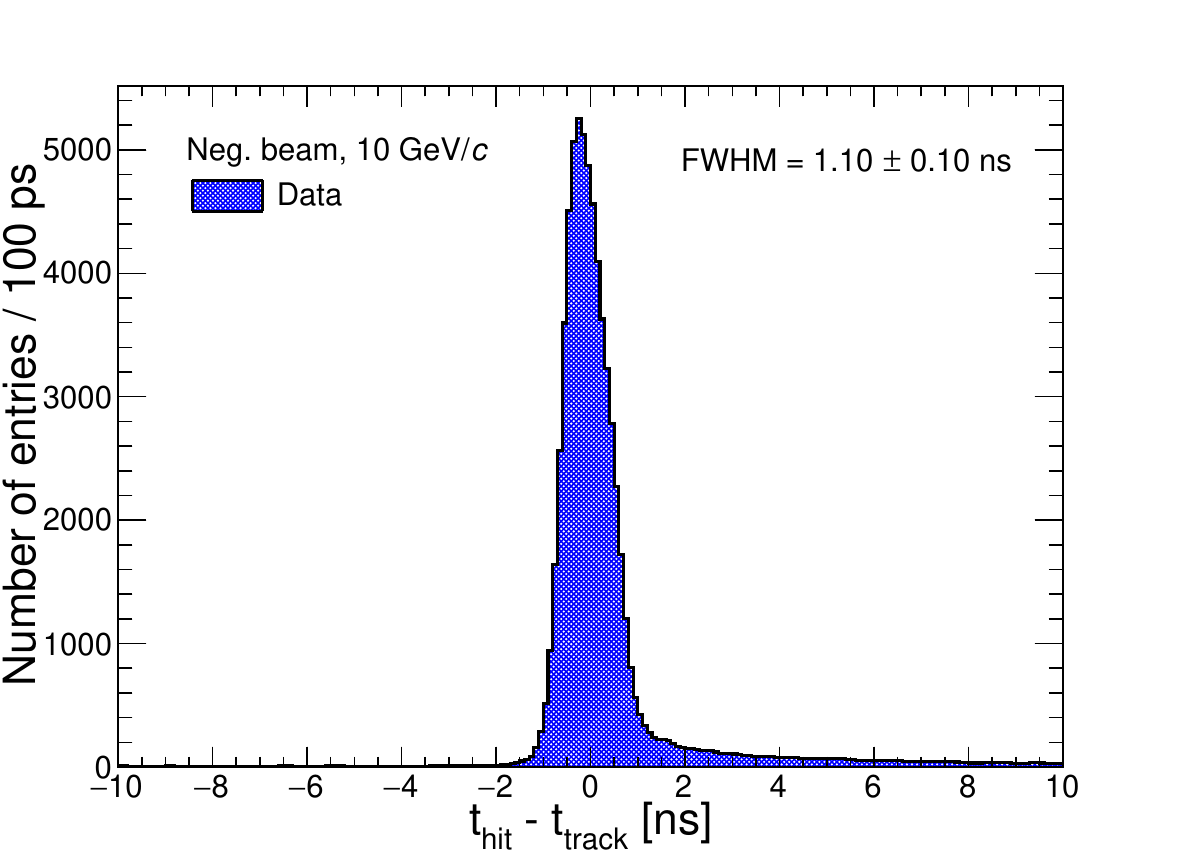} 
\includegraphics[width=0.95\columnwidth]{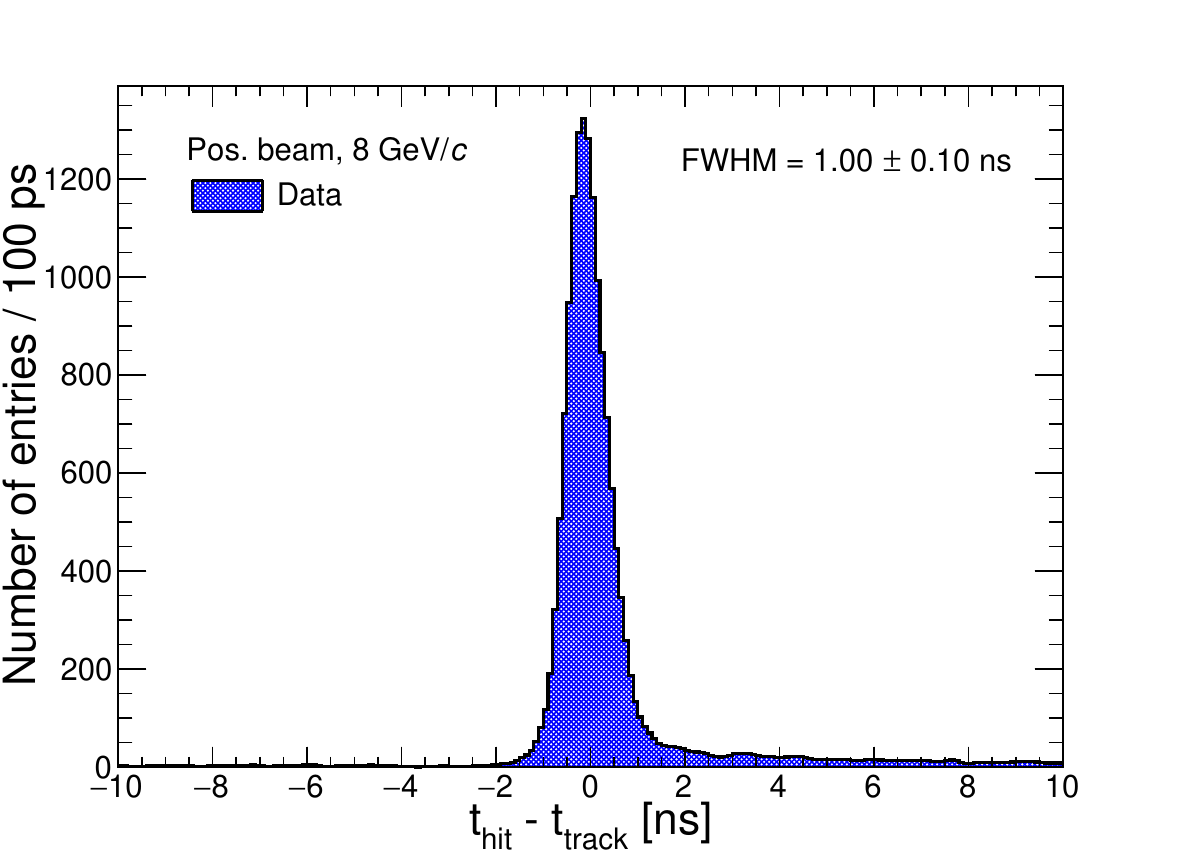} 
\caption{Distribution of the time differences between the array cell firing time $\text{t}_{\text{hit}}$ and the firing time $\text{t}_{\text{track}}$ of the central array charged particle cluster cell with maximum number of PEs in the events with the negative charged beam at 10 GeV/$c$ momentum (top panel) and with the positive charged beam at 8 GeV/$c$ momentum (bottom panel).} 
\label{fig:timing_single}
\end{figure}

\begin{figure*}[!h!t]
\centering
\begin{overpic}[width=0.95\columnwidth,tics=10]{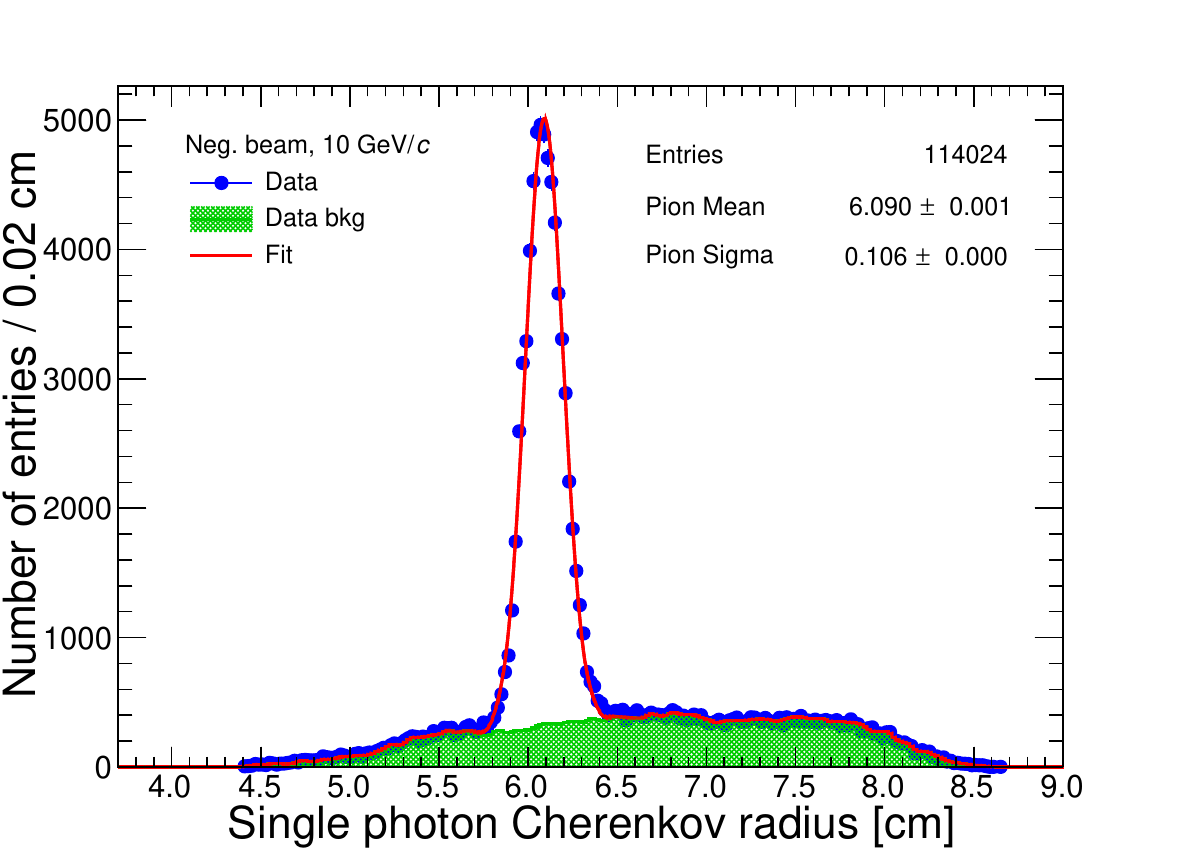}\put(15,43){\scriptsize All hits}\end{overpic}
\begin{overpic}[width=0.95\columnwidth,tics=10]{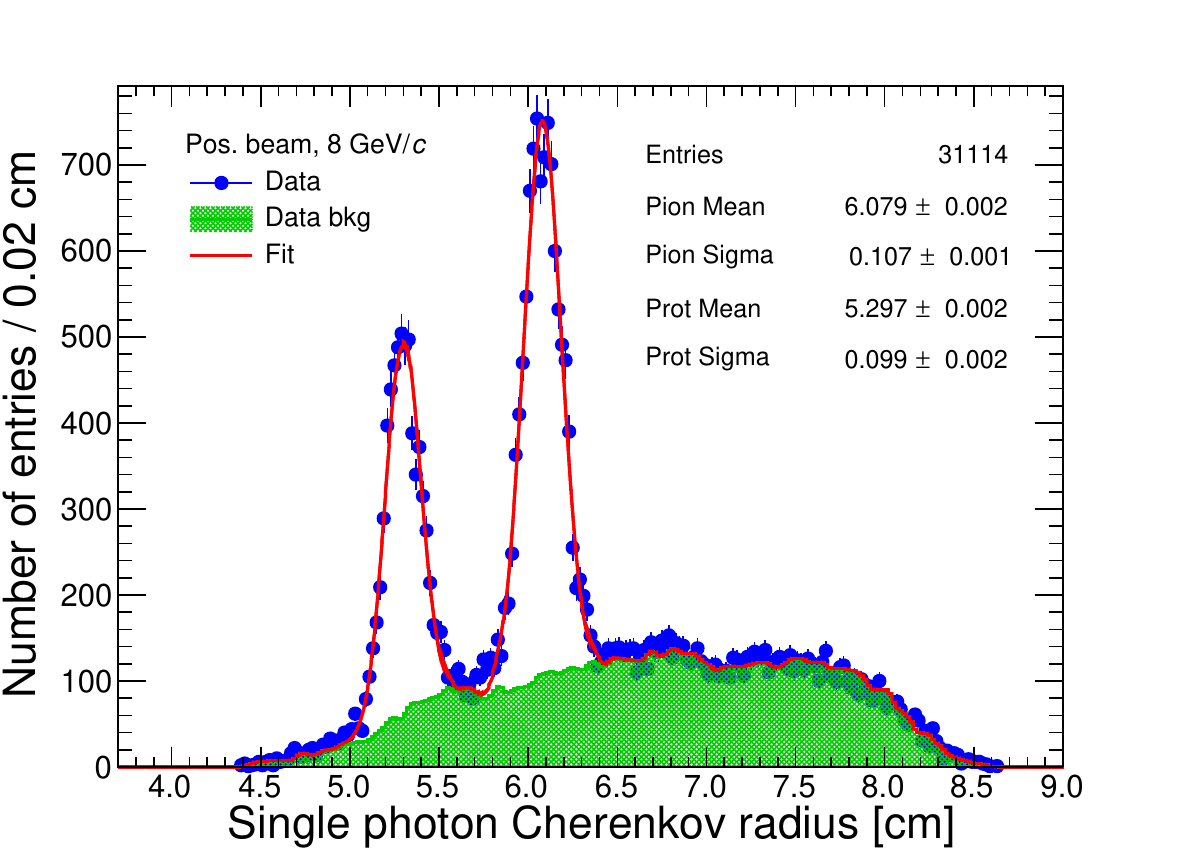}\put(15,43){\scriptsize All hits}\end{overpic}
\begin{overpic}[width=0.95\columnwidth,tics=10]{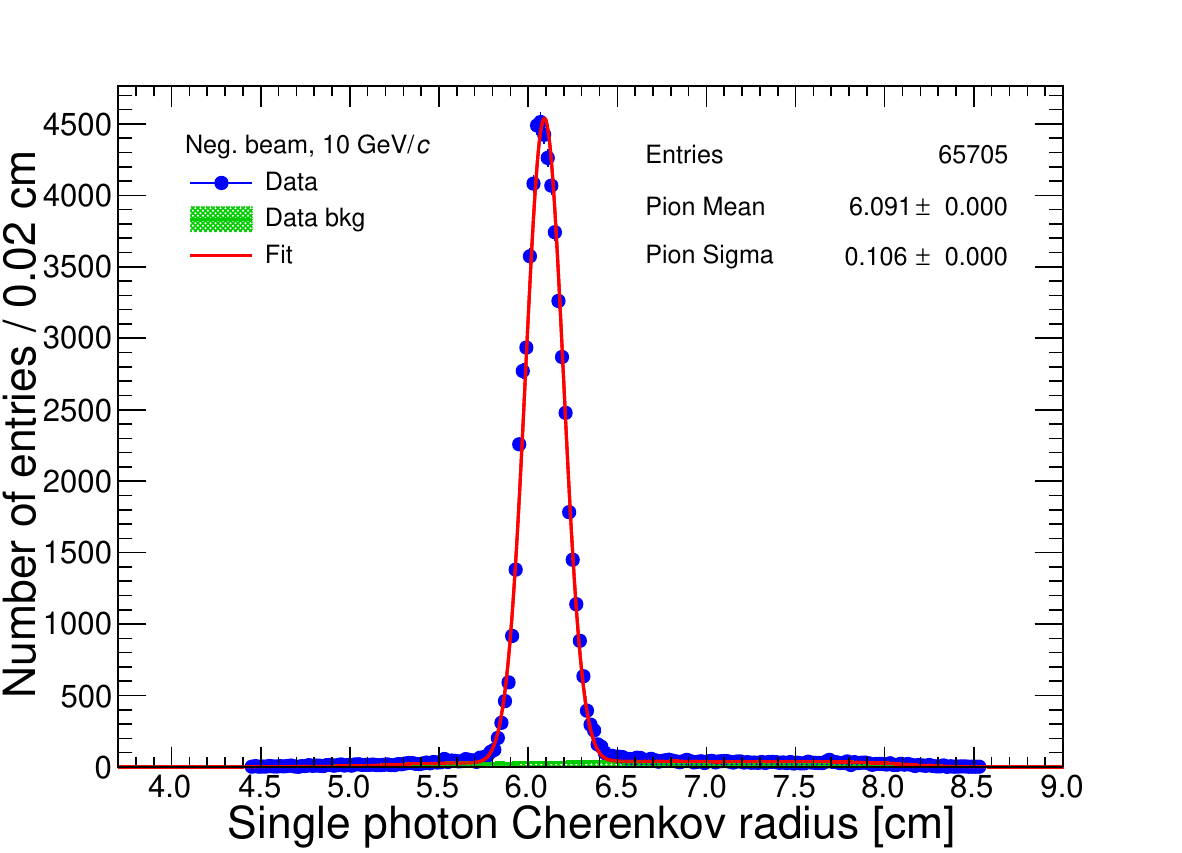}\put(15,43){\scriptsize $\pm$5 ns cut}\end{overpic}
\begin{overpic}[width=0.95\columnwidth,tics=10]{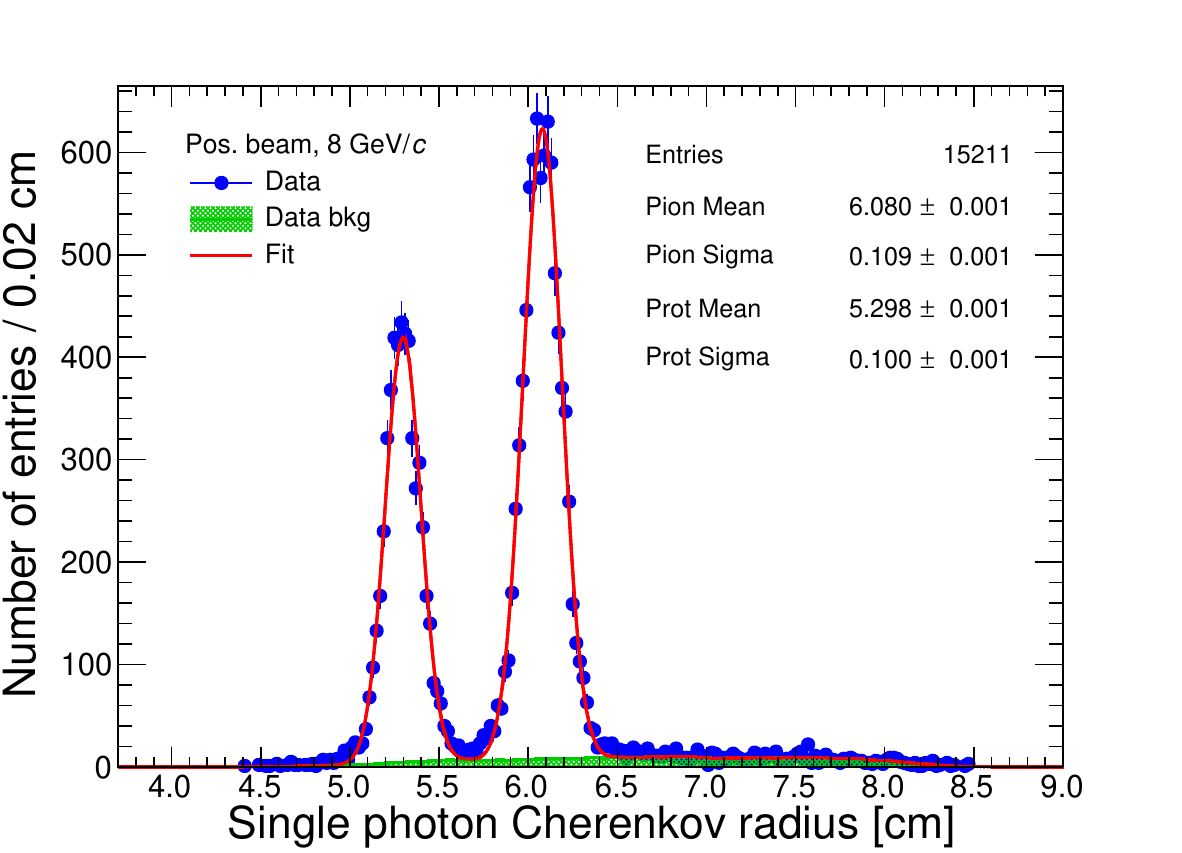}\put(15,43){\scriptsize $\pm$5 ns cut}\end{overpic}
\caption{Distribution of the reconstructed aerogel single photon Cherenkov radius for the negative charged beam at 10 GeV/$c$ momentum (left column) and for the positive charged beam at 8 GeV/$c$ momentum (right column). The bottom plots show the distribution requiring hit-track matching in a $\pm 5$ ns time window. The
vertical bars represent the statistical uncertainties. The fit with the sum of a Gaussian for the pion peak (or two Gaussians for the pion and proton peaks) and our template background distribution is also shown. }
\label{fig:radius_single}
\end{figure*}

\begin{figure*}[!h!t]
\centering
\begin{overpic}[width=0.95\columnwidth,tics=10]{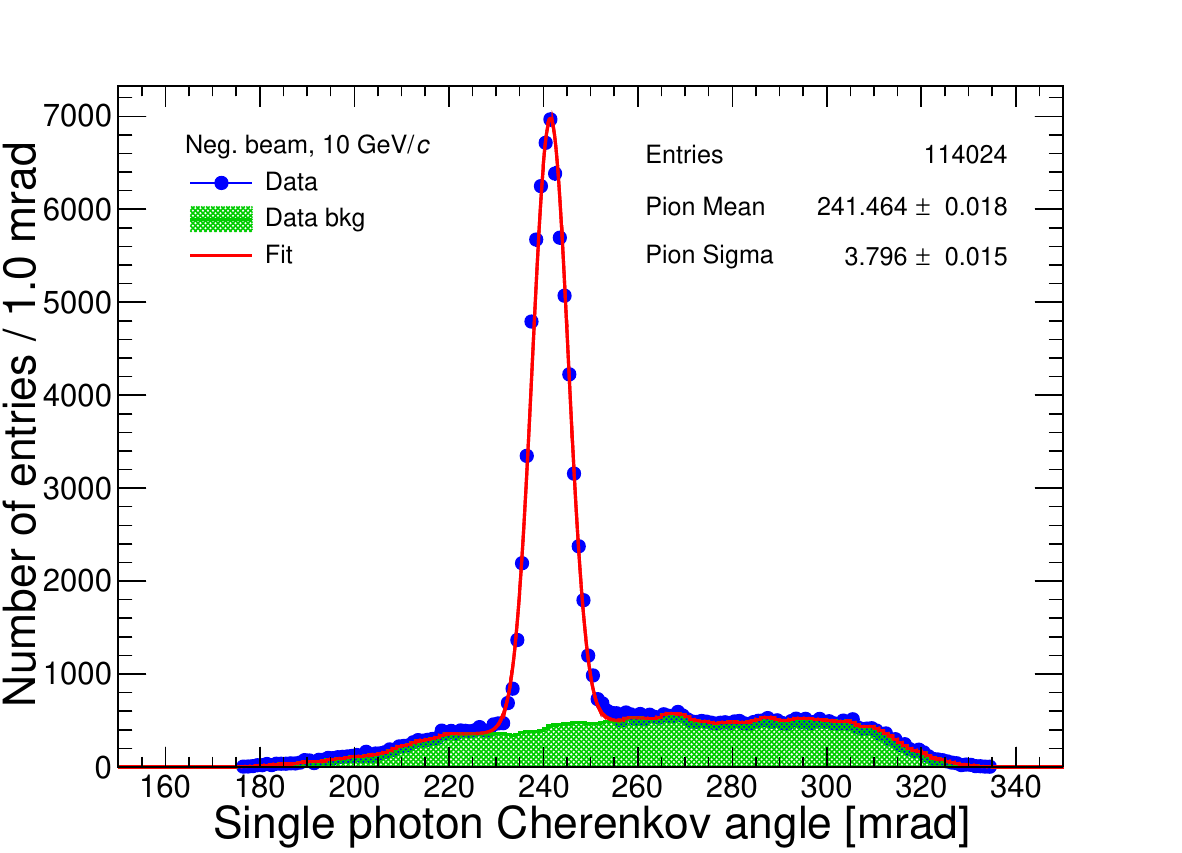}\put(15,43){\scriptsize All hits}\end{overpic}
\begin{overpic}[width=0.95\columnwidth,tics=10]{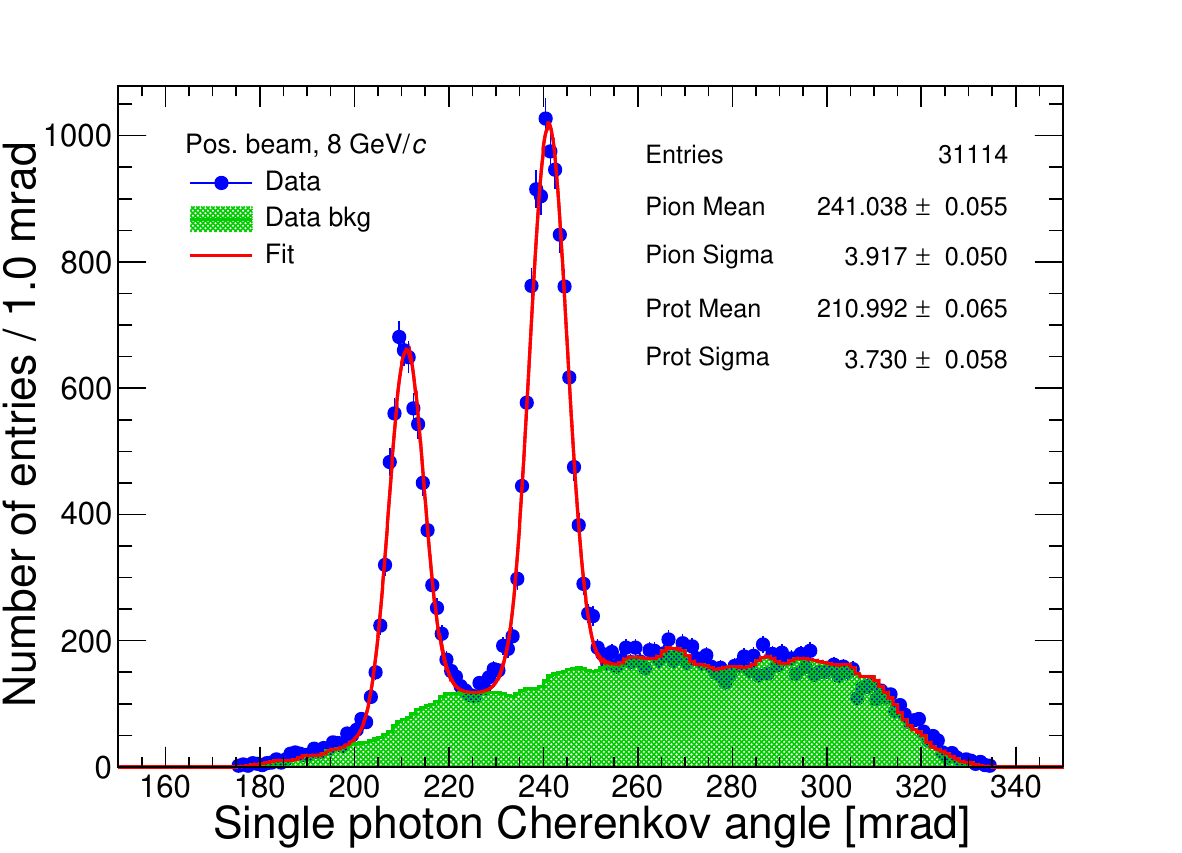}\put(15,43){\scriptsize All hits}\end{overpic}
\begin{overpic}[width=0.95\columnwidth,tics=10]{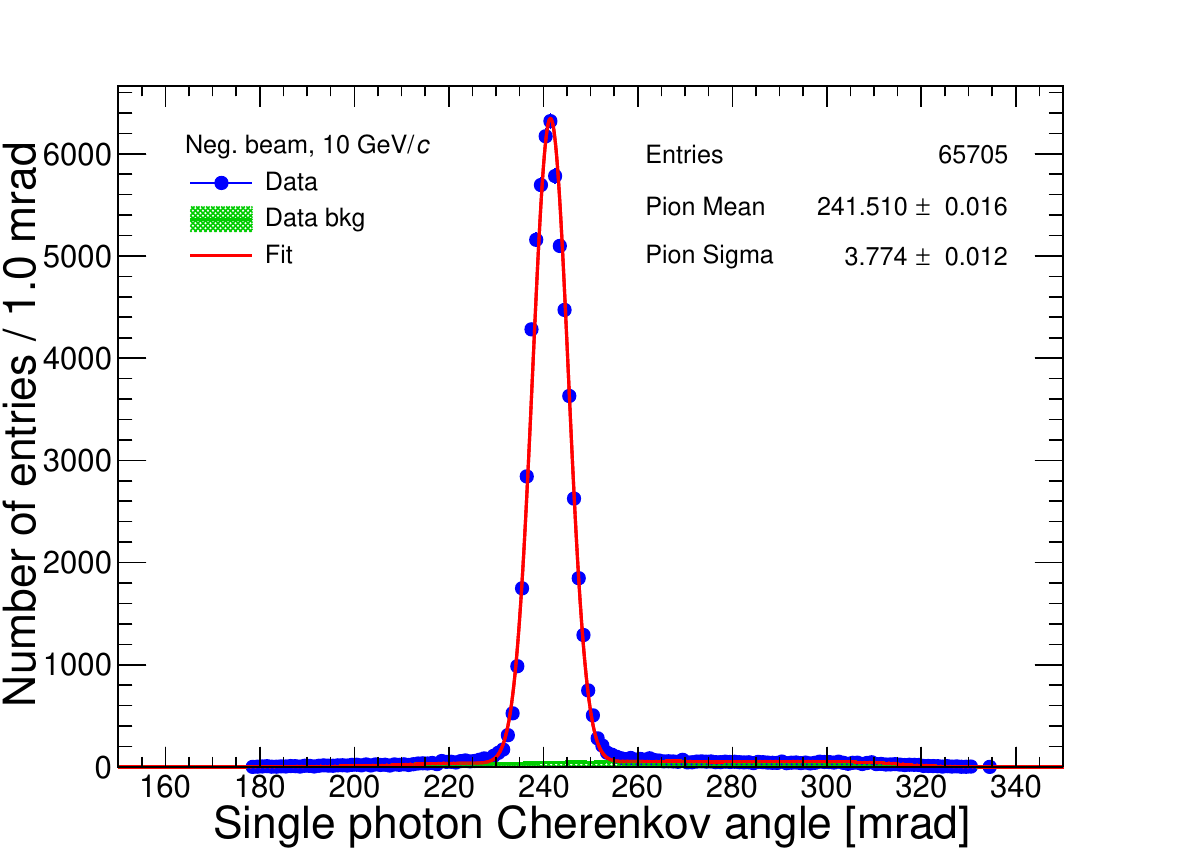}\put(15,43){\scriptsize $\pm$5 ns cut}\end{overpic}
\begin{overpic}[width=0.95\columnwidth,tics=10]{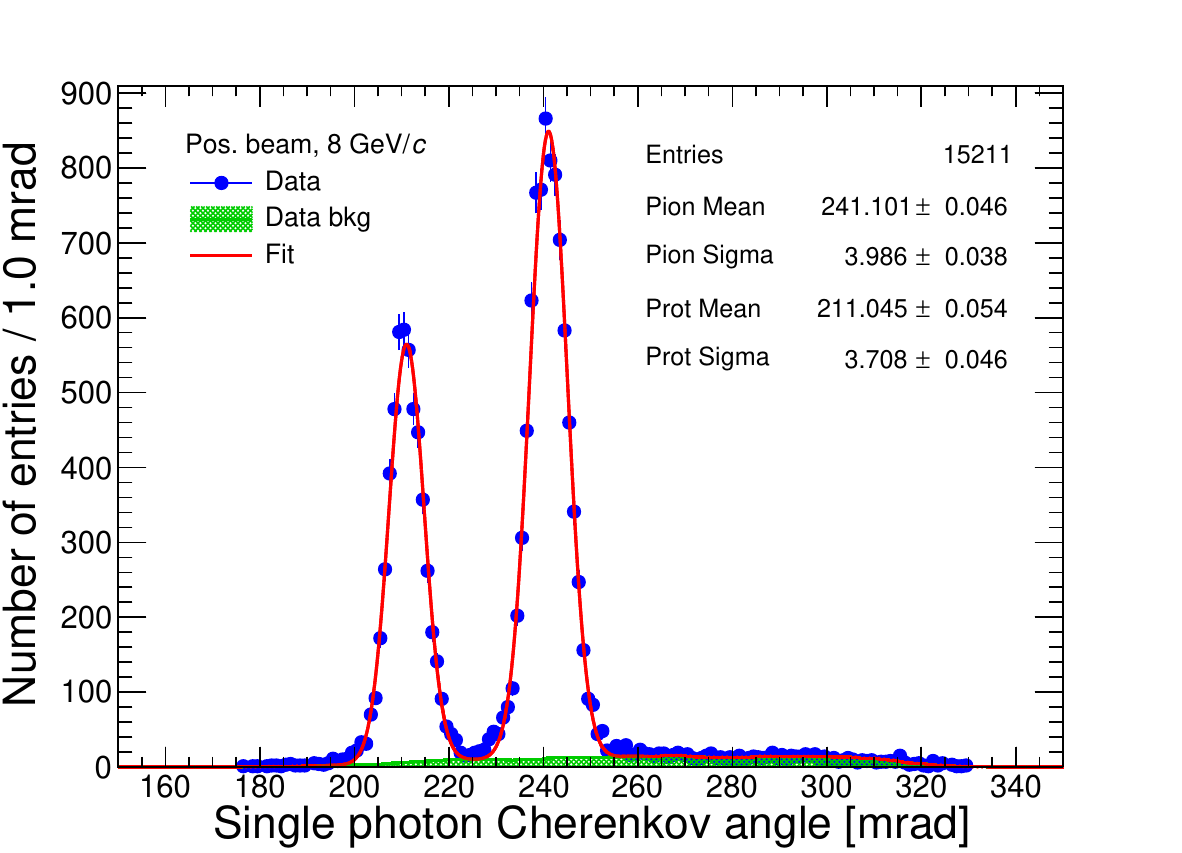}\put(15,43){\scriptsize $\pm$5 ns cut}\end{overpic}
\caption{Distribution of the reconstructed aerogel single photon Cherenkov angle for the negative charged beam at 10 GeV/$c$ momentum (left column) and for the positive charged beam at 8 GeV/$c$ momentum (right column). The bottom plots show the distribution requiring hit-track matching in a $\pm 5$ ns time window. The
vertical bars represent the statistical uncertainties. The fit with the sum of a Gaussian for the pion peak (or two Gaussians for the pion and proton peaks) and our template background distribution is also shown. }
\label{fig:theta_single}
\end{figure*}

\begin{figure}[!t] 
\centerline{\begin{overpic}[width=0.95\columnwidth,tics=10]{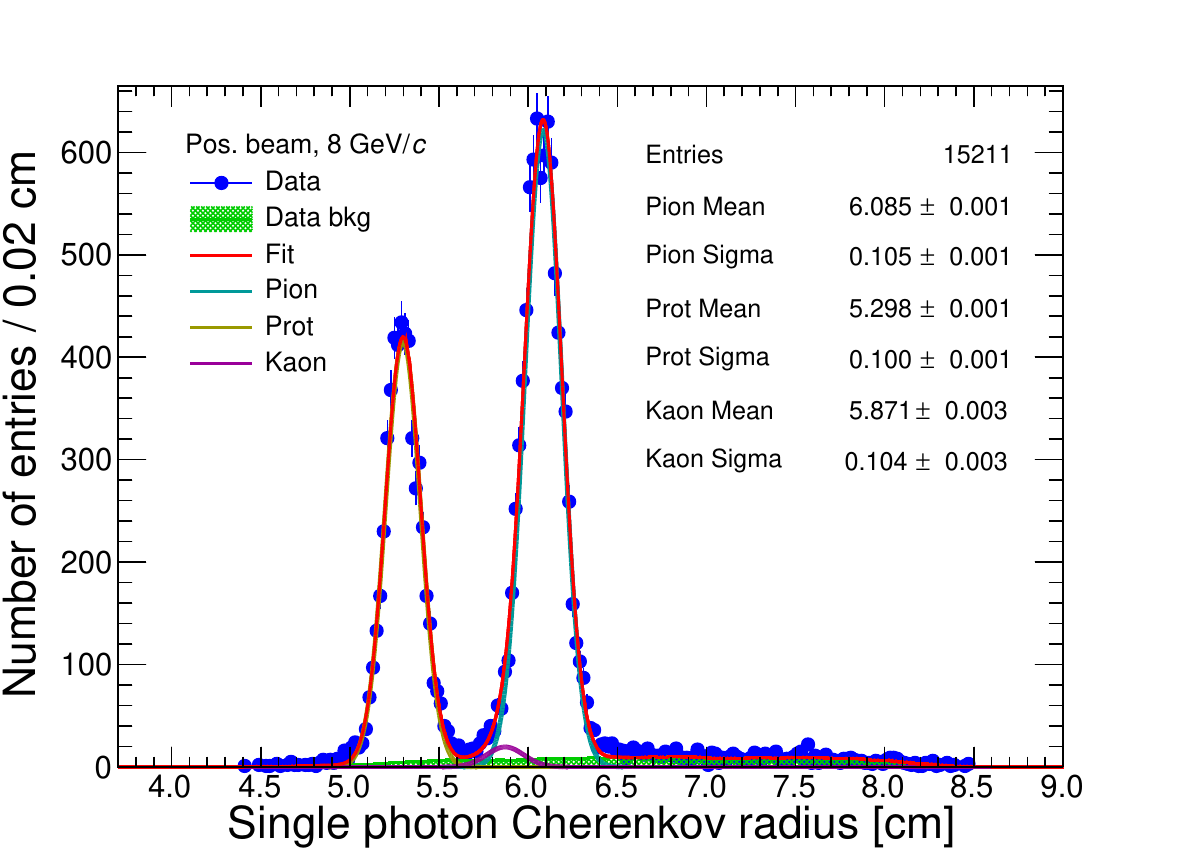}\put(15,33){\scriptsize $\pm$5 ns cut}\end{overpic}}
\centerline{\begin{overpic}[width=0.95\columnwidth,tics=10]{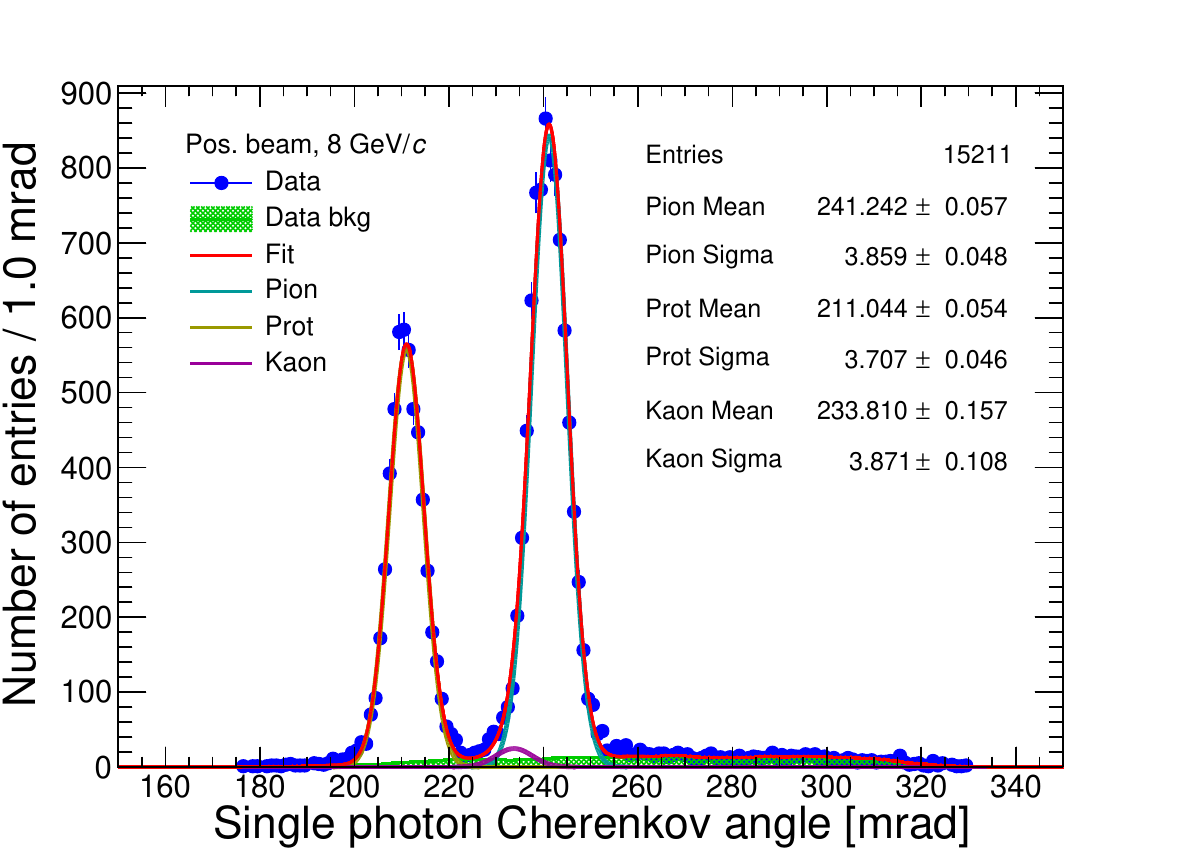}\put(15,33){\scriptsize $\pm$5 ns cut}\end{overpic}}
\caption{Distribution of the reconstructed aerogel single photon Cherenkov radius (top) and angle (bottom) for the positive charged beam at 8 GeV/$c$ momentum requiring hit-track matching in a $\pm 5$ ns time window. The
vertical bars represent the statistical uncertainties. The fit with three Gaussians for pion, proton and kaon signals and our template background distribution is also shown. }
\label{fig:theta_single_plus8GeV_cut_v3_triple}
\end{figure}
 
Cherenkov photons from aerogel were searched for within the hits in the ring arrays.
The hit Z coordinates of the ring arrays were assumed to the nominal Z coordinates of the SiPM arrays while the X-Y positions were calculated according the nominal fired channel positions.

We performed Cherenkov angle studies assuming Z-photon emission coordinate in the median plane of the aerogel tile normal to the beam direction, corresponding to a 1 cm depth in the aerogel tile.
The corresponding X-Y position of the photon emission point was calculated according to the extrapolated charged track parameters. The resulting uncertainty on both the X and Y position of emission point was evaluated to be about 300 $\mu$m.

The reconstructed spatial distributions of the extrapolated X-Y emission points in aerogel and of the hits in the ring arrays for the negative charged beam at 10 GeV/$\it{c}$ and the positive charged beam at 8 GeV/$\it{c}$ are shown in Fig.~\ref{fig:hit_maps}. The expected clusters due to Cherenkov photons from pions and protons are clearly visible. The corresponding Cherenkov rings centered in the most probable X-Y emission point are also shown.

Various sources of both correlated and uncorrelated background hits affect the distributions.
Correlated background counts mainly come from Rayleigh scattering of Cherenkov photons emitted in the near-ultraviolet region in aerogel \cite{Altamura:2024jp}.
The differential Rayleigh scattering cross section of low-energy photons is proportional to 1 + $\cos^2{\theta_{\text{sc}}}$, where $\theta_{\text{sc}}$ is the scattering angle relative to the initial photon direction \cite{RayleighAngularDistributionPolarized,RayleighAngularDistribution}.
Correlated background also accounts for photons by secondaries, such as delta-electrons or products from primary particle interactions, as well as the contribution of both direct and Rayleigh-scattered photons undergoing multiple reflections at the interfaces between the materials in the cylinder before being detected. 
As a first approximation, this results in almost uniformly distributed background hits in each of the ring arrays.

Uncorrelated background counts mainly come from the random firing of ring array channels due to the SiPM DCR and electronics noise.
In order to model the uncorrelated background hits, dedicated runs were performed by acquiring events in which a 10 $\mu$s delay was added to the external trigger signal. 
This approach allowed us to achieve an excellent modeling of noisy channels and channel-by-channel fluctuations.
Due to the similarities in the topology of correlated and uncorrelated background, we assumed the same spatial distributions extrapolated for uncorrelated background to model the correlated background contribution as well, up to a multiplicative factor. 
Therefore, the overall background model considered for the radius and angle distributions in the following is:

\begin{equation}
    \frac{dN_{\text{bkg}}}{d\text{X}} =  k \left( \frac{dN_{\text{bkg}}}{d\text{X}} \right)_\text{template},  
\end{equation}
where $X$ is the Cherenkov radius or angle, $k$ is a free parameter and $(dN_{\text{bkg}} / d\text{X})_\text{template}$ is the corresponding template distribution resulting from the spatial distribution of background hits evaluated as discussed above.

For consistency check, we performed an alternative background extrapolation directly from signal runs (i.e. without accounting any delay in the external trigger signal) considering the relative timing between the ring array cell firing time $\text{t}_{\text{hit}}$ and the firing time $\text{t}_{\text{track}}$ of the central array charged particle cluster cell with the maximum number of PEs in the event. 
Examples of the resulting relative timing distributions for the negative charged beam at 10 GeV/$c$ and the positive charged beam at 8 GeV/$c$ are shown in Fig. \ref{fig:timing_single}. 
Signal hits are clustered in the Gaussian core of the distribution, having a sigma of approximately 400~ps. 
This value is dominated by the large time jitter of the Petiroc 2A ASICs for the single PE signals in the ring arrays.
Conversely, a contribution smaller than 100~ps was measured for the central array in which there are  charged particle signals with many PEs due to the prompt Cherenkov photons emitted in the fused silica window~\cite{Mazziotta_IWORID}.  
The right tail of the distributions in Fig. \ref{fig:timing_single} corresponds to delayed cross-talk and afterpulsing following a true photon signal in channels where the single PE amplitude is lower than the acquisition threshold, but the subsequent pile-up of delayed signals in the same SiPM allows threshold crossing. 
Therefore, the corresponding spatial distribution of uncorrelated background hits has been extracted by selecting only hits such that $|\text{t}_{\text{hit}} - \text{t}_{\text{track}}|~>~50$~ns. 
The resulting distributions are consistent with the ones obtained from the runs with a 10 $\mu$s delay in the external trigger signal.

The relative timing was also used for the uncorrelated background suppression. Uncorrelated background filtering was performed by selecting only hits such that $|\text{t}_{\text{\text{hit}}} - \text{t}_{\text{track}}| <~\text{X}$~ns.
The impact of the considered time window width on the signal efficiency and the signal-to-background ratio leading to the choice of an optimal value of X is discussed in the following.

We performed the Cherenkov angle reconstruction through a simple geometric backpropagation from the hit positions in the ring arrays to the extrapolated emission point in the radiator.
In the backpropagation procedure, corrections by Snell’s law were applied to account for the refraction at the interfaces between different media assuming the corresponding refractive indices at 400 nm wavelength. 
 We assumed all the observed hits in the ring arrays as candidate Cherenkov photons emitted in aerogel by the considered charged track.

\begin{figure}[t] 
\centering
\includegraphics[width=0.95\columnwidth]{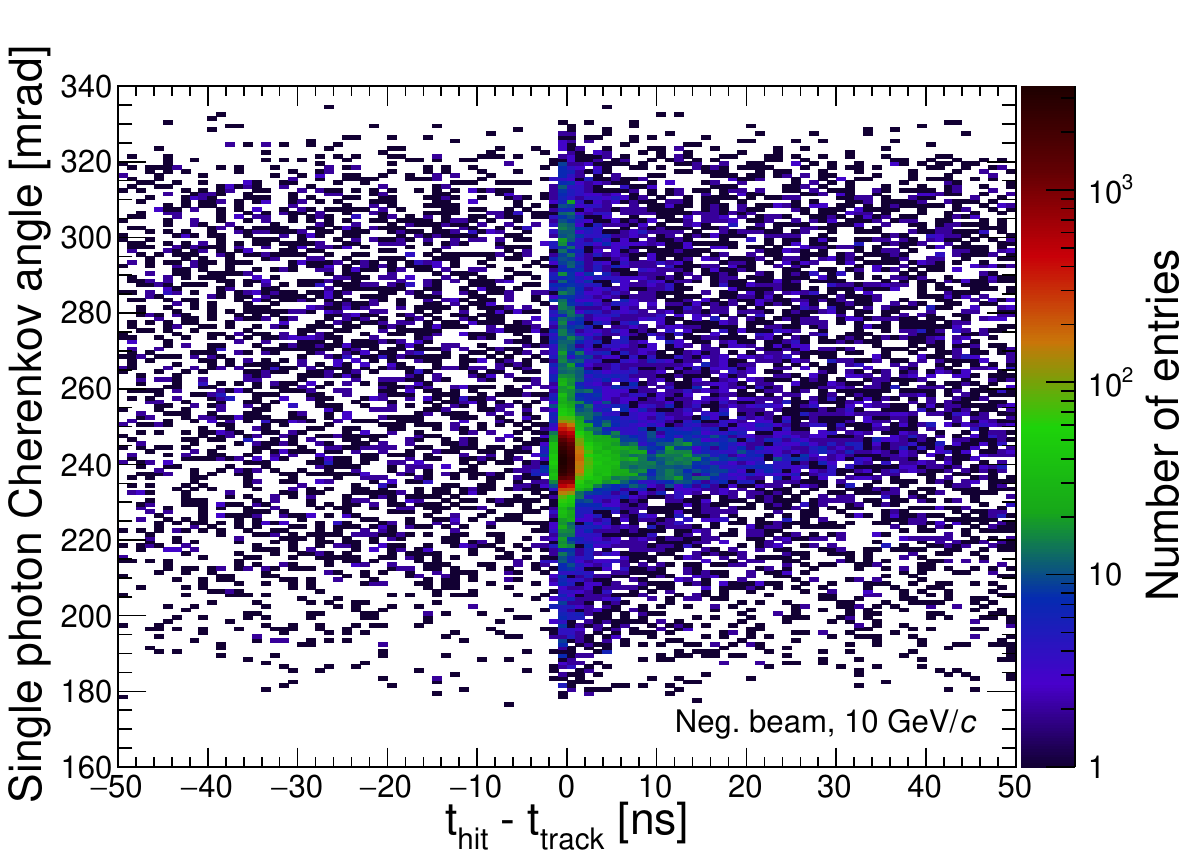}
\includegraphics[width=0.95\columnwidth]{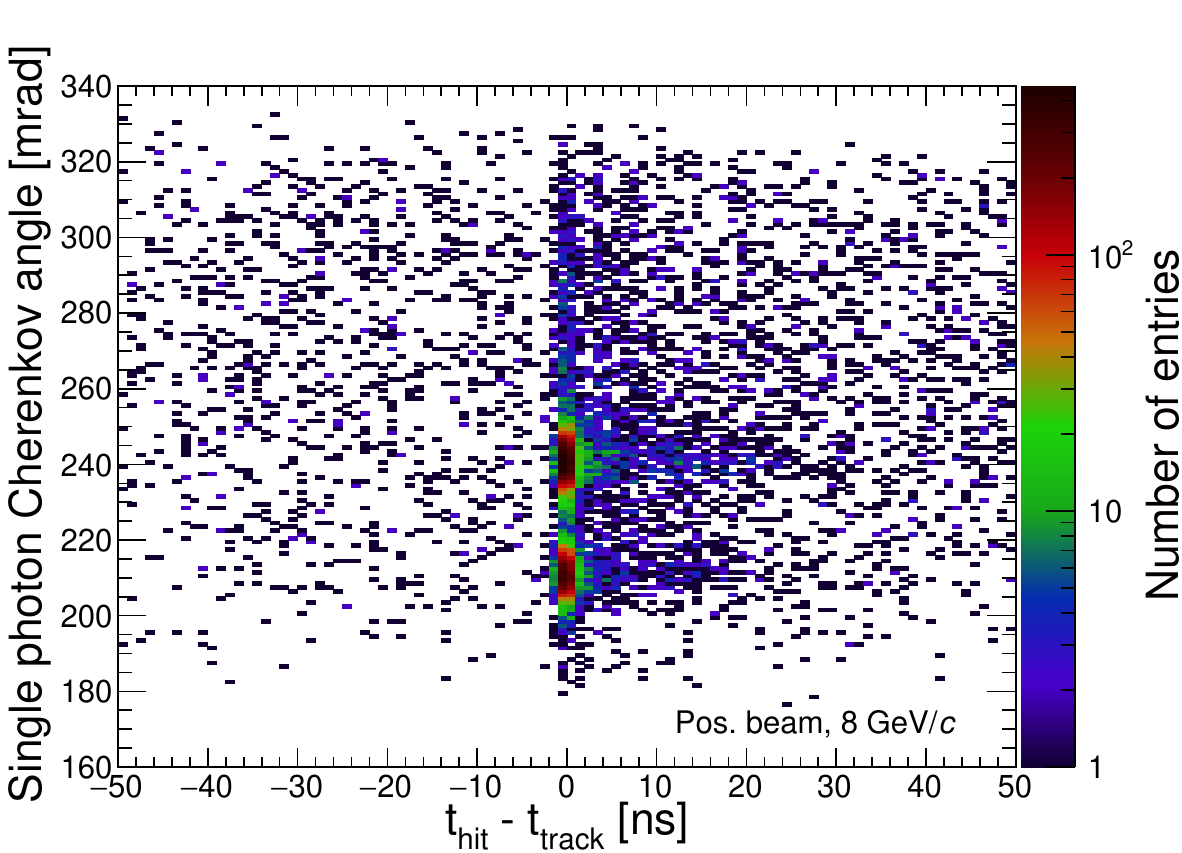}
\caption{
Distribution of the reconstructed aerogel single photon Cherenkov angle as a function of the relative timing between the ring arrays and the central array for the negative charged beam at 10 GeV/$c$ momentum (top panel) and for the positive charged beam at 8 GeV/$c$ momentum (bottom panel).}
\label{fig:angle_vs_time_2D}
\end{figure}

\begin{figure*}[!h!t]
\centering
\includegraphics[width=0.95\columnwidth]{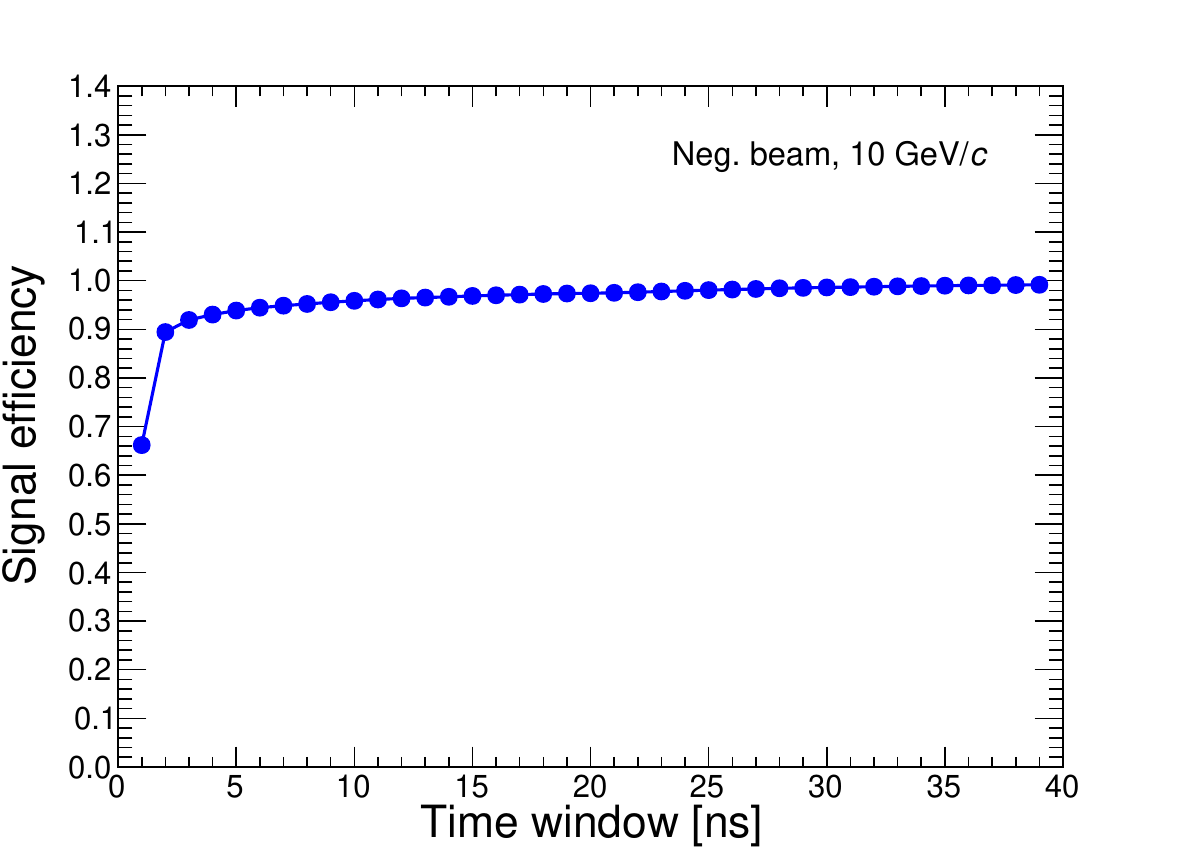}
\includegraphics[width=0.95\columnwidth]{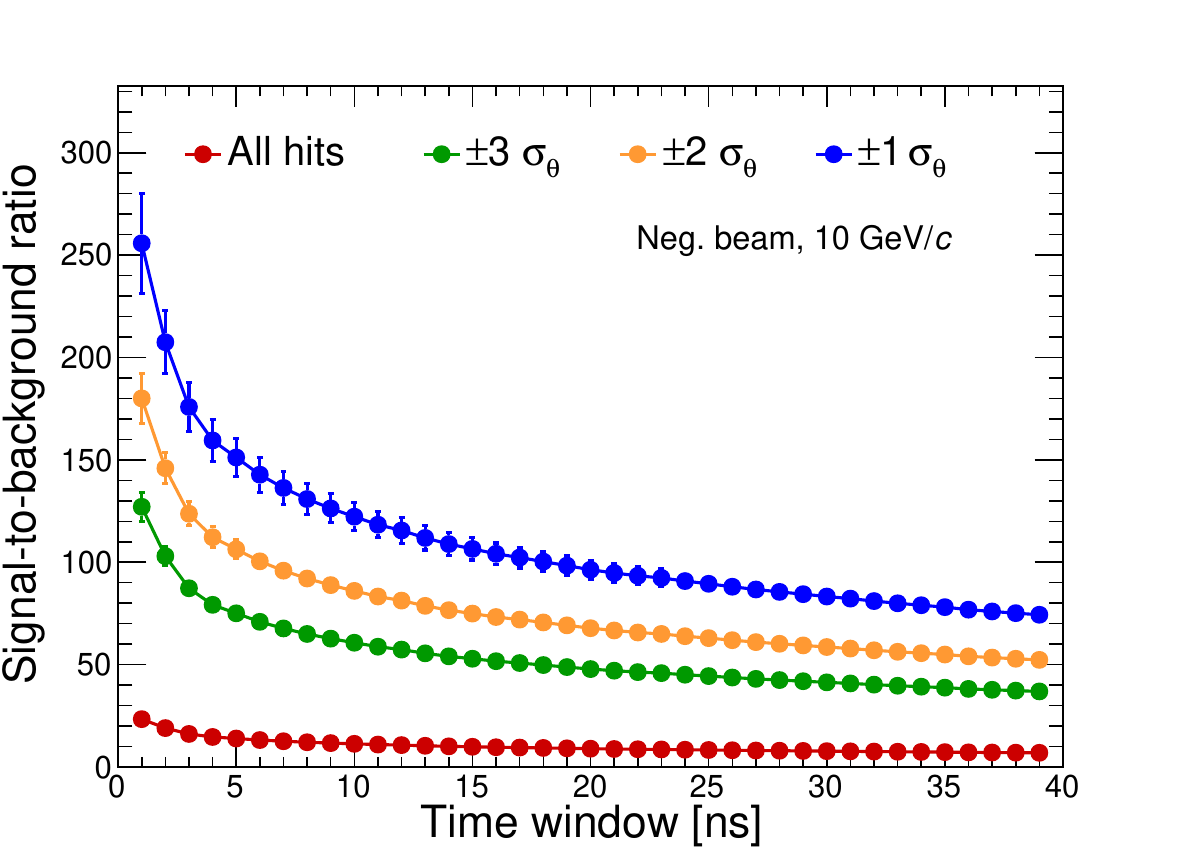}
\includegraphics[width=0.95\columnwidth]{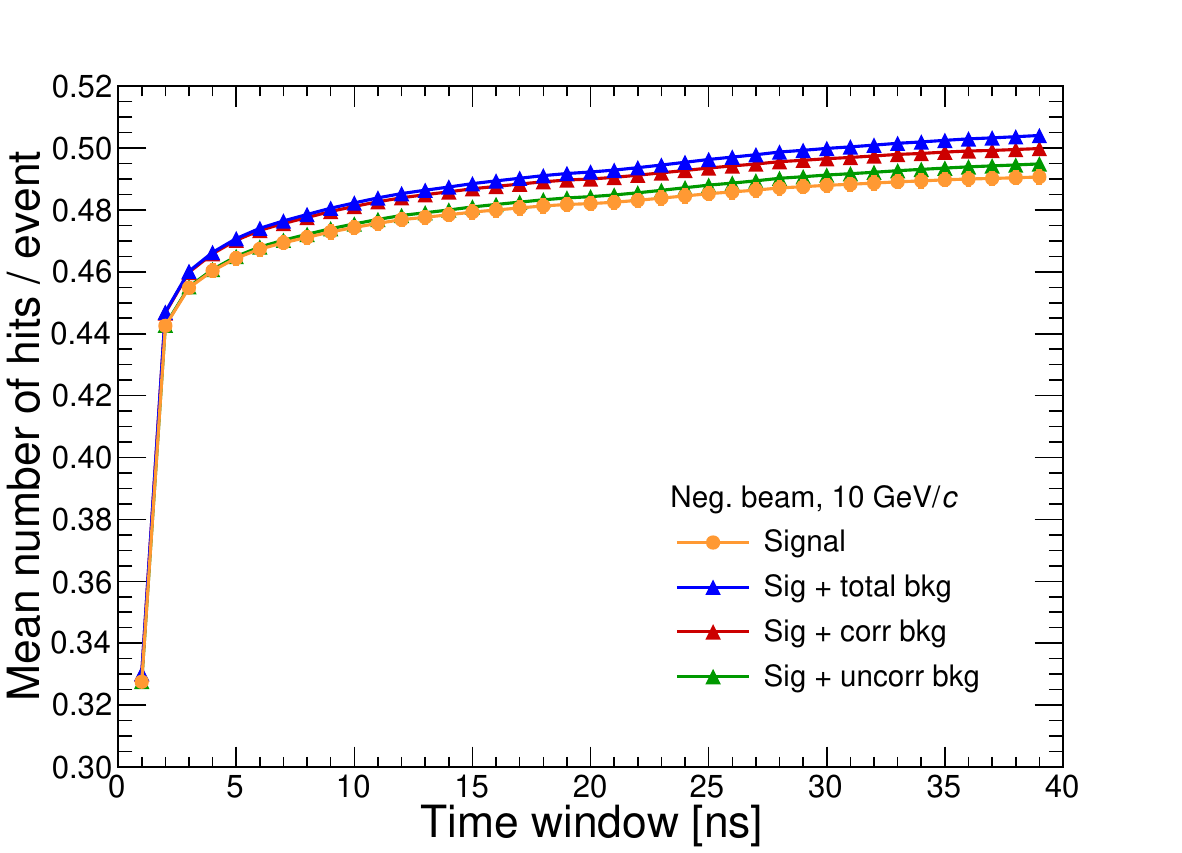}
\includegraphics[width=0.95\columnwidth]{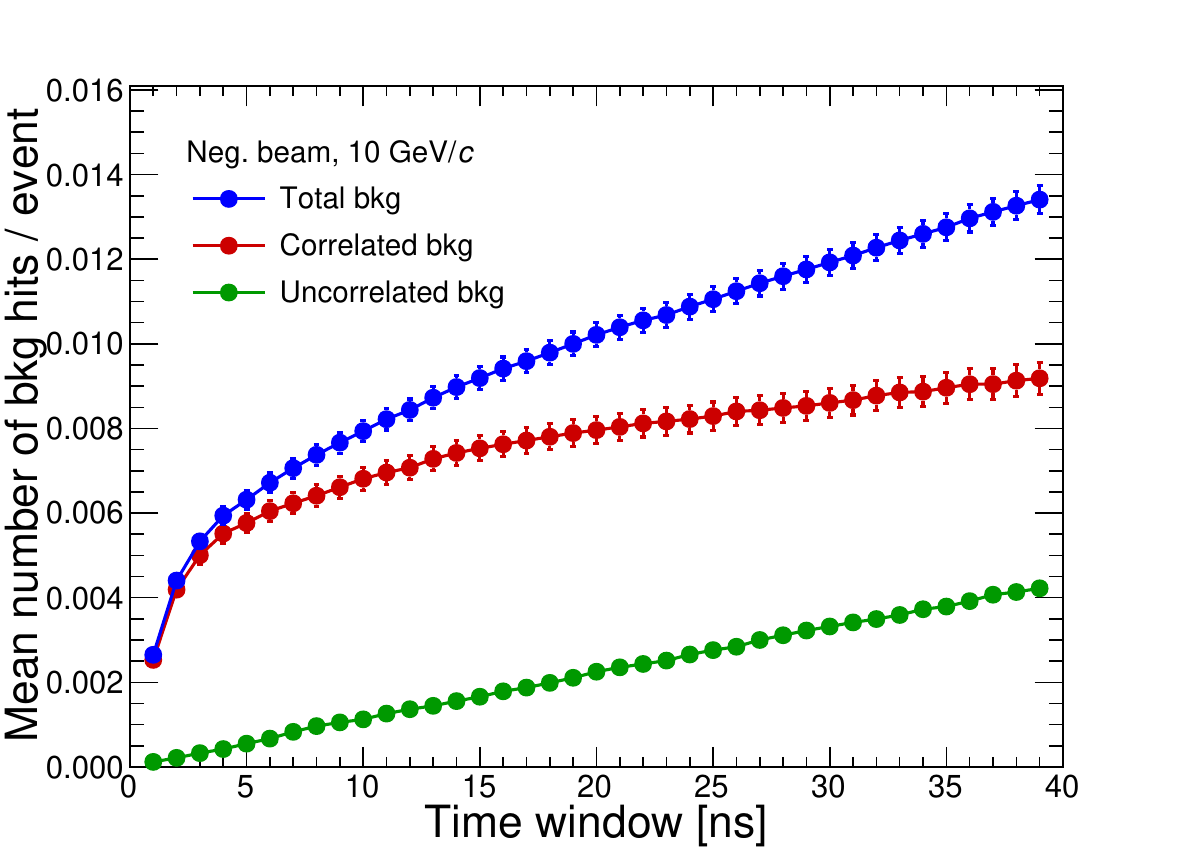}
\caption{Signal efficiency (top left), signal-to-background ratio (top right) and extrapolated mean number of detected signal hits, correlated background hits, uncorrelated background hits and total background hits per track (bottom left and bottom right) as a function of the full width of the relative hit-track timing window centered on $\text{t}_{\text{track}}$ for the negative charged beam at 10 GeV/$c$. The signal-to-background ratio is evaluated considering the hits in both full acceptance and intervals with measured angle $\theta_{\text{hit}}$ such that $|\theta_{\text{hit}}  - \theta_c|<\text{N}\sigma_c$, with $\text{N}=$ 1, 2 and 3. The mean number of signal and background hits per event is evaluated in the interval $|\theta_{\text{hit}}  - \theta_c|<3\sigma_c$.}
\label{fig:vs_time_window_hit_track}
\end{figure*}

The distributions of the reconstructed Cherenkov ring radius and  emission angle at single hit level obtained considering all available ring array hits for the negative charged beam at 10 GeV/$\it{c}$ and the positive charged beam at 8 GeV/$\it{c}$ are shown in the top plots of Fig.~\ref{fig:radius_single} and Fig.~\ref{fig:theta_single}.
The corresponding distributions filtered by selecting only hits such that $|\text{t}_{\text{hit}} - \text{t}_{\text{track}}| < 5$ ns are shown in the bottom plots of the same figures.
The expected peaks due to Cherenkov photons from pions and protons are clearly visible. The distributions are fitted with the sum of Gaussian distribution corresponding to the signal peaks and a template background distribution, as discussed above. 

The mean value of the Gaussian distribution provides the most probable Cherenkov ring radius $R_c$ and emission angle $\theta_c$, while the sigma represents the single photon radial resolution $\sigma_R$ and angular resolution $\sigma_c$. 
The observed values of $R_c$ for pions at 10 GeV/c, pions at 8 GeV/c and protons at 8 GeV/c momentum are 6.09, 6.08 and 5.30 cm, respectively. The corresponding values for $\theta_c$ are 241.5 mrad, 241.1 mrad and 211.1 mrad, respectively. 
The measured values are consistent with the theoretical expected scaling of the Cherenkov ring radius and emission angle as a function of both the particle mass and momentum.  

Although the expected fraction of protons in the CERN-PS positive beam at 8 GeV/$c$ momentum is larger than the one of pions \cite{PS_beam_composition_buona}, we observe a smaller proton peak with respect to the one of pions. The observed relative fractions have been verified to be consistent with expectations accounting both the lower Cherenkov photon yield of protons in aerogel, proportional to  sin$^{2}\theta_c$ \cite{Eugenio_RICH}, and the limited ring array acceptance to proton ring, having a smaller radius with respect to the pion one as outlined in Fig. \ref{fig:hit_maps}.

We measured single hit values of $\sigma_R$ and $\sigma_c$ of approximately 0.1 cm and 3.8 mrad, respectively, for both the 10 GeV/$c$ pion peak and the 8 GeV/$c$ proton peak.
The pion resolutions at 10 GeV/$c$ and 8 GeV/$c$ are expected to be consistent.
However, a $\sigma_c$ worsening by a few tenths of a mrad is observed for the 8 GeV/$c$ pion peak with respect to the corresponding peak at 10 GeV/$c$. 
In addition, a slight discrepancy between data and the fit model is observed on the left tail of the pion peak in the angular and radial distributions at 8 GeV/$c$ momentum.
These effects are due to the contribution of kaons, having an absolute beam fraction of approximately 3\%.
The expected $\theta_c$ of kaons at 8 GeV/$c$ momentum is approximately 234 mrad. 
The corresponding expected $\sigma_c$ is comparable with the one of other species.
Therefore, the contribution of kaons to the radial and angular distributions in Figs.~\ref{fig:radius_single} and ~\ref{fig:theta_single} falls on the left tail of the pion peak, resulting in a larger apparent $\sigma_R$ and $\sigma_c$.
Fig.~\ref{fig:theta_single_plus8GeV_cut_v3_triple} shows the same background-filtered radial and angular distributions for the positive charged beam at 8 GeV/$c$ momentum shown in Figs.~\ref{fig:radius_single} and ~\ref{fig:theta_single} fitted by introducing an additional Gaussian in the fit model.
As expected, the $\sigma_R$ and $\sigma_c$ of the pion Gaussian are reduced and are consistent with the value at 10 GeV/$c$ within the uncertainties.
Although kaons are also present at 10 GeV/$c$, the much smaller  yield and the closer $\theta_c$ relative to pions make their contribution not possible to be disentangled at single hit level. The same consideration applies to the lower than 2\% expected electron and positron component of the negative charged beam at 10 GeV/$c$ and the positive charged beam at 8 GeV/$c$, respectively. 

\begin{figure}[!ht] 
\centering
\includegraphics[width=0.95\columnwidth]{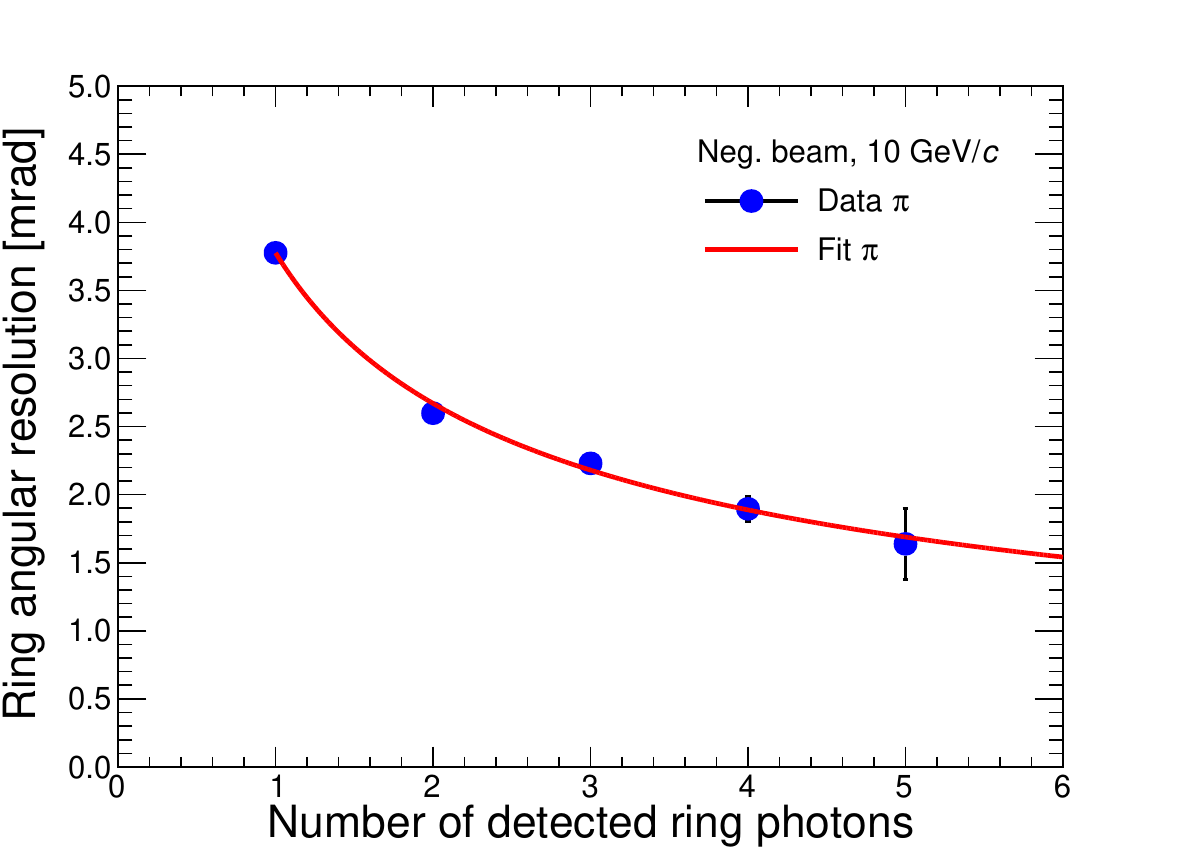}
\includegraphics[width=0.95\columnwidth]{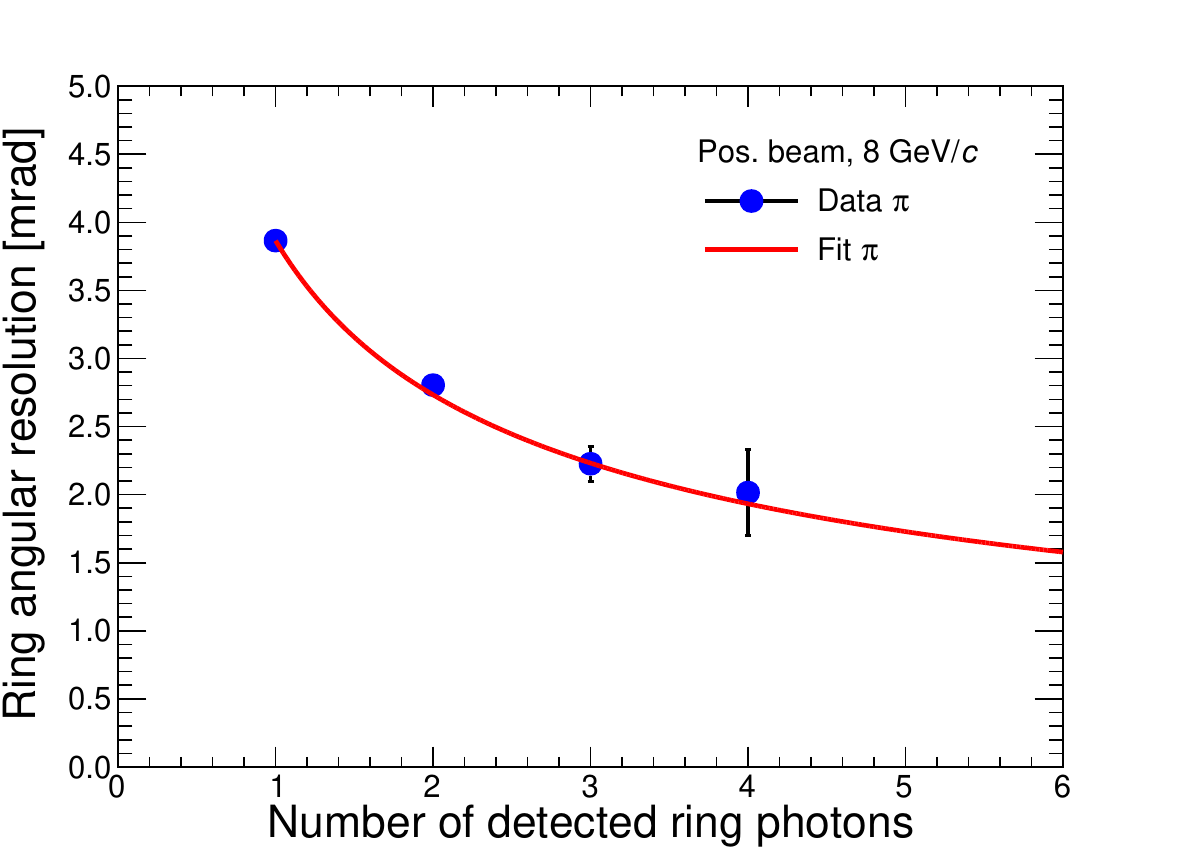}
\caption{Cherenkov angular resolution as a function of the number $N_{\text{ph}}$ of detected photons in the pion ring for the negative charged beam at 10 GeV/$c$ momentum (top panel) and for the positive charged beam at 8 GeV/$c$ momentum (bottom panel). The vertical bars represent the statistical uncertainties. The fit with the expected 1/$\sqrt{N_{\text{ph}}}$ scaling relative to the single photon angular resolution is also shown.}
\label{fig:theta_ring}
\end{figure}

\begin{figure}[!ht] 
\centering
\includegraphics[width=0.95\columnwidth]{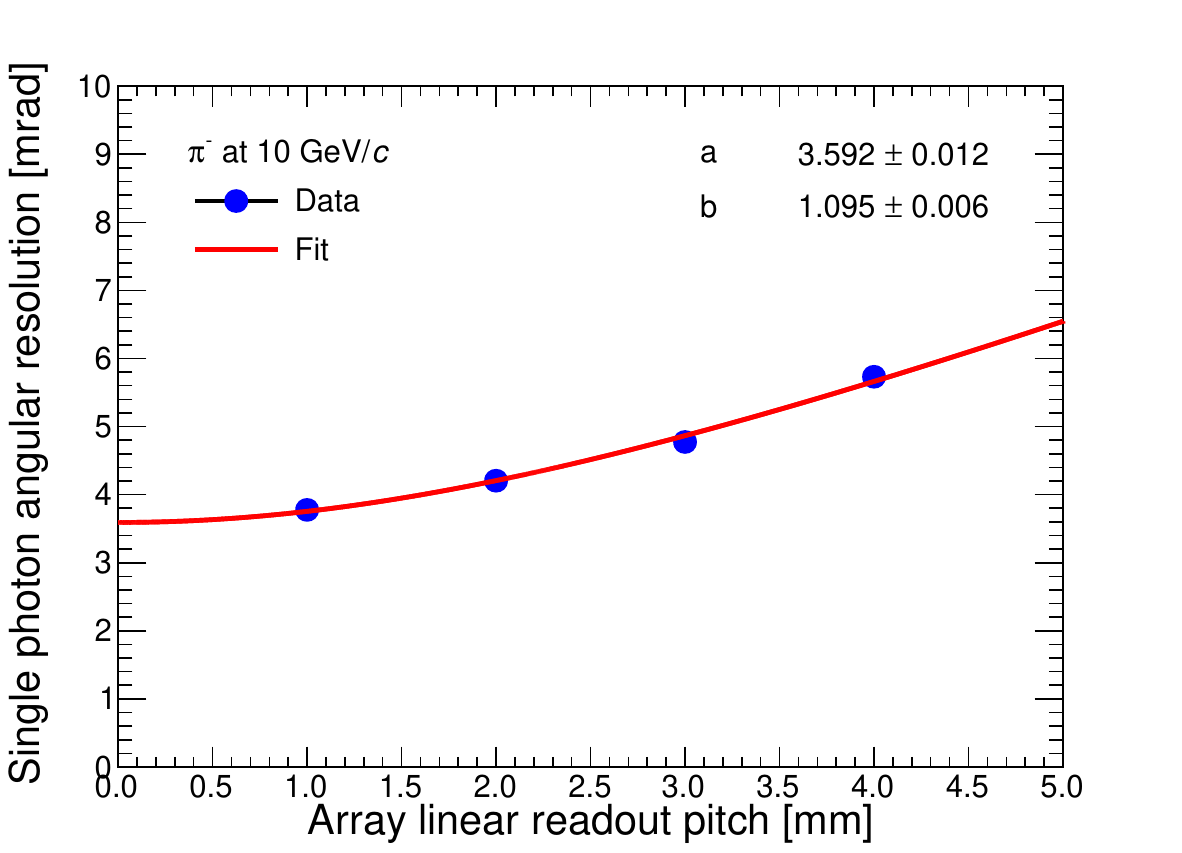}
\includegraphics[width=0.95\columnwidth]{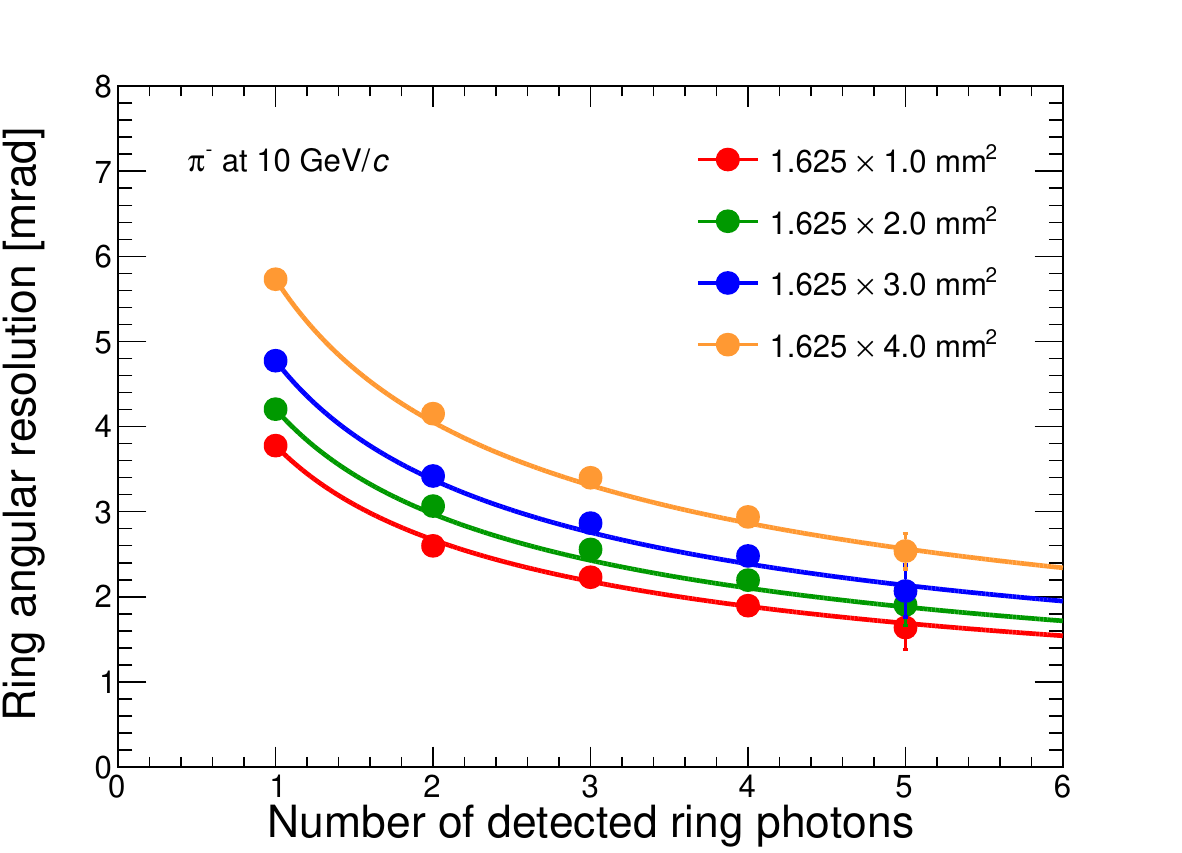}
\caption{
Extrapolated aerogel single photon angular resolution (top panel) and ring angular resolution as a function of the number $N_{\text{ph}}$ of detected ring photons (bottom panel) for increasing readout pitch by grouping at software level adjacent ring array $1\!\times\!1.625$~mm$^2$ readout pixels for negative pions at 10 GeV/$c$ momentum. The vertical bars represent the statistical uncertainties. The fits with Eq. \ref{Eq:sigma_single_vs_pixel_size} and the scaling of the ring angular resolution as 1/$\sqrt{N_{\text{ph}}}$ are also shown.}
\label{fig:sigma_vs_pixel_size}
\end{figure}

Considering all the available hits in the ring arrays, approximately 45\% of those contributing to the distributions in Figs.~\ref{fig:radius_single}~and~\ref{fig:theta_single} comes from the cumulative uncorrelated background sampled in the full 550 ns acquisition window.
Excellent background suppression is achieved using the information on timing.
With the considered $|\text{t}_{\text{hit}} - \text{t}_{\text{track}}| < 5$ ns cut, the candidate Cherenkov photon fraction of background hits decreases from 45\% down to approximately 8\% at the cost of a less than 5\% reduction in the signal photon yield.
The majority of the surviving background hits is due to correlated background, such as Cherenkov photons undergoing Rayleigh scattering before leaving aerogel.
Correlated background hits in the ring arrays are synchronous with the hits due to unscattered signal photons and cannot be discarded based on timing.
The effect is visible in Fig. \ref{fig:angle_vs_time_2D}, showing the 2D angular distributions of the reconstructed Cherenkov angle as a function of the relative hit-track timing. The sparse uniformly distributed background corresponds to uncorrelated background due to the electronics noise and DCR signals. The clusters at the expected Cherenkov angles for the different particle species correspond to direct Cherenkov photons. The vertical band superimposed to the clusters corresponds to correlated background.
The horizontal bands at the expected Cherenkov angles are due to discriminator threshold effects, as discussed for the right tail of the distributions in Fig. \ref{fig:timing_single}.

Fig. \ref{fig:vs_time_window_hit_track} shows the signal efficiency and the signal-to-background ratio considering the hits in both full acceptance and intervals with measured angle $\theta_{\text{hit}}$ such that $|\theta_{\text{hit}}  - \theta_c|<\text{N}\sigma_c$, with $\text{N}=$ 1, 2 and 3, as a function of the full width of the relative hit-track timing window centered on $\text{t}_{\text{track}}$ for the negative charged beam at 10 GeV/$c$.
The signal efficiency was calculated as the ratio of the area under the fit Gaussian peak for the considered time window to the area under the fit Gaussian peak considering all hits. Values above 95\% are achieved considering a time window larger than 10 ns, corresponding to the $|\text{t}_{\text{hit}} - \text{t}_{\text{track}}| < 5$ ns cut considered for the distributions in Figs. \ref{fig:radius_single}, \ref{fig:theta_single} and \ref{fig:theta_single_plus8GeV_cut_v3_triple}.
The signal-to-background ratio was calculated as the ratio of the area under the fit Gaussian peak 
to the area under the fit background model.
Values above 55, 90 and 130 are achieved at 3, 2 and 1 $\sigma_c$, respectively, considering with a time window shorter than 10 ns.

\begin{figure}[t] 
\centering
\includegraphics[width=0.95\columnwidth]{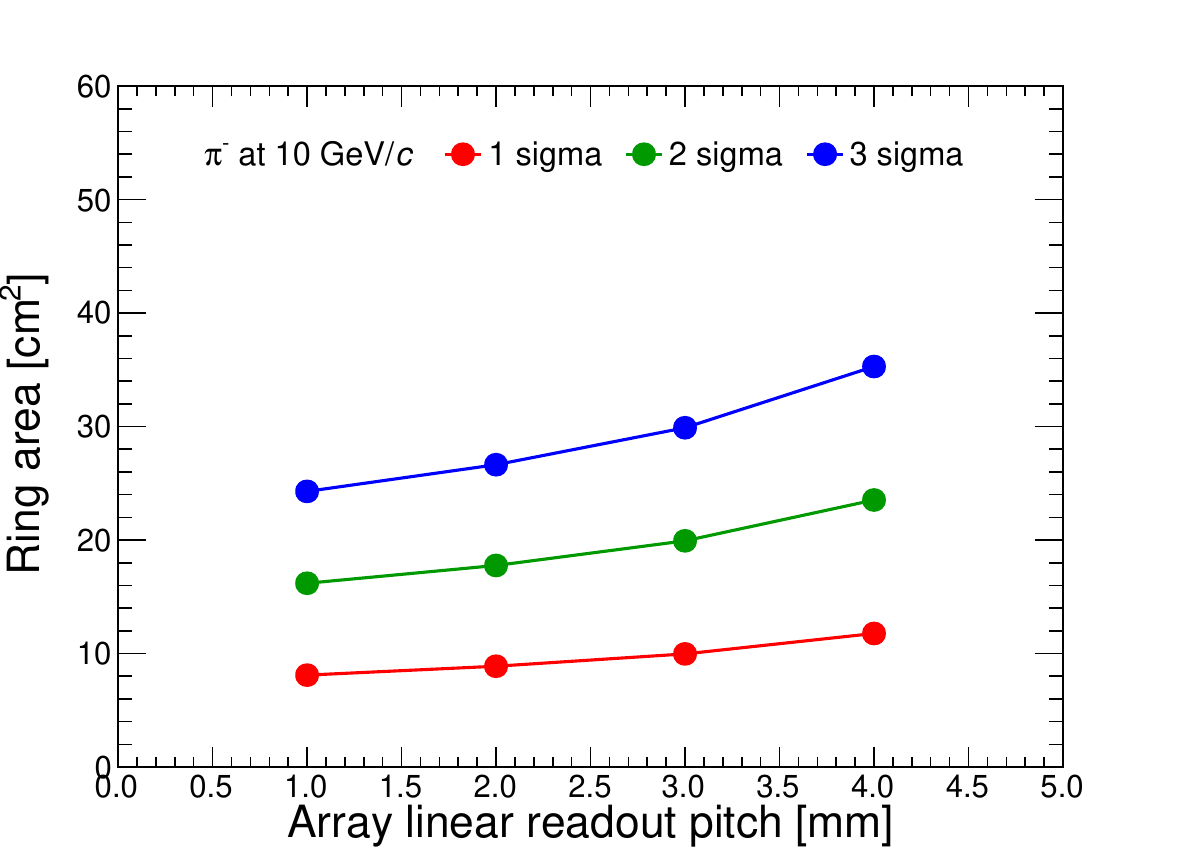}
\includegraphics[width=0.95\columnwidth]{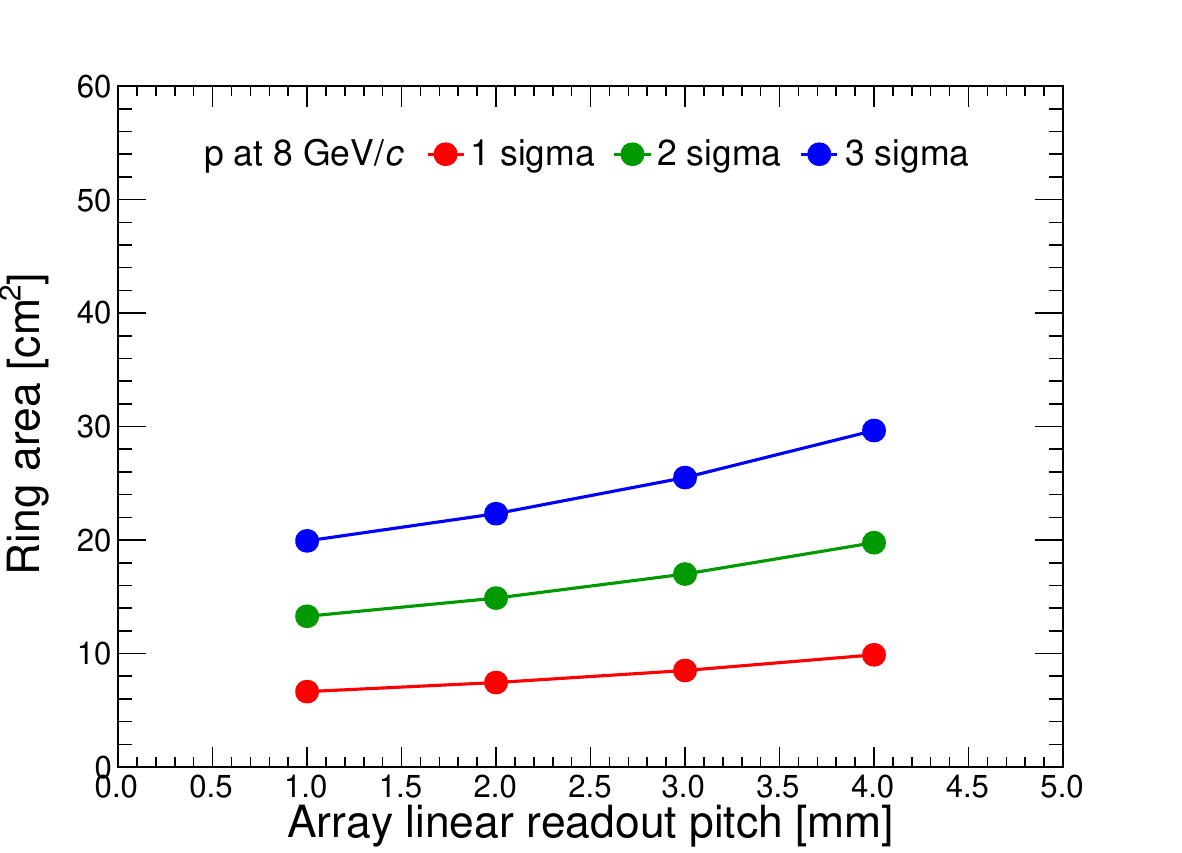}
\caption{
Extrapolated aerogel ring fiducial area where Cherenkov photons are clustered assuming full ring coverage with SiPMs for increasing linear readout pitch by grouping at software level adjacent ring array $1\!\times\!1.625$~mm$^2$ readout pixels for pions at 10 GeV/$c$ momentum (top panel) and protons at 8 GeV/$c$ momentum (bottom panel).}
\label{fig:area_ring_full_coverage}
\end{figure}

Fig \ref{fig:vs_time_window_hit_track} also shows the extrapolated mean number of detected signal hits, correlated background hits, uncorrelated background hits and total background hits per track with reconstructed angle $\theta_{\text{hit}}$ such that $|\theta_{\text{hit}}  - \theta_c|<3\sigma_c$  as a function of the full width of the relative hit-track timing window for the negative charged beam at 10 GeV/$c$. 
The number of signal hits and correlated background hits rapidly increases as the time window enlarges, approaching saturation in a few ns, modulo threshold effects as discussed above. On the contrary, the number of uncorrelated background hits increases linearly with the time window. However, it remains lower than the amount of correlated background hits in the whole timing interval considered for this analysis. 
The abundance of correlated background hits has been verified to be consistent with the measured scattering length of the tested aerogel tile~\cite{Altamura:2024jp}.
We measured on average more than 0.47 signal hits on top of an overall background of less than 0.008 hits per track for the considered 10 ns window. 
 
We also measured the expected improvement of the angular resolution as $ \propto 1/\sqrt{\text{N}_{\text{ph}}}$ with the increasing number $\text{N}_{\text{ph}}$ of detected photons.
Despite the limited ring acceptance, corresponding to approximately 1/30 of the full ring, the large accumulated statistics was sufficient to extend the analysis up to $\text{N}_{\text{ph}} =$ 3, 4 or 5 ring photons depending on the dataset and the considered particle species.
For this study, the ring Cherenkov angle associated with a given track was determined by averaging all the reconstructed angles within the interval $[~\theta_c~-~3\sigma_c,~\theta_c~+~3\sigma_c~]$ such that $|\text{t}_{\text{hit}} - \text{t}_{\text{track}}| < 5$ ns. Here, $\theta_c$ and $\sigma_c$ refer to the most probable Cherenkov emission angle and single photon angular resolution corresponding to the considered particle species and momentum.
The resulting ring angular resolution as a function of $N_{\text{ph}}$ for the negative charged beam at 10 GeV/$\it{c}$ and the positive charged beam at 8 GeV/$\it{c}$ momentum is shown in Fig. \ref{fig:theta_ring}. The value of the ring angular resolution for a given $\text{N}_{\text{ph}}$ are extracted as the sigma of the Gaussian fit to the corresponding ring Charenkov angle distributions.
The available points are fitted with the expected $1/\sqrt{\text{N}_{\text{ph}}}$ scaling relative to the single photon angular resolution.
We extrapolate a ring angular resolution better than 1.5 mrad for $\text{N}_{\text{ph}}>6$. 
For a full-scale system with the photosensitive surface fully covered by SiPMs the average $N_{\text{ph}}$ is expected to be about 28, 
also correcting for inefficiencies due to the operation threshold of the arrays~\cite{Nicassio_proceeding_IWASI}.
Therefore, we predict an achievable $e/\pi$, $\pi/K$, and $K/p$ separation better than 3$\sigma$ up to 3, 10 and 17 GeV/c momentum, respectively.
These results validate the ALICE~3 bRICH design with respect to the resolution expectations as outlined in Sec.~\ref{sec:intro} considering brand-new SiPMs.

The results discussed so far refer to a readout of each of the eight S13552 ring arrays corresponding to 32 $1\!\times\!1.625$~mm$^2$ pixels, obtained by grouping four adjacent strips having linear pitch of 0.25 mm.
However, for an effective design of the full-scale system, 
the optimization of the cell size is crucial.
In particular, a larger pixel size may be required to reduce the number of electronics channels and keep limited the system complexity, provided that the angular resolution still matches the target PID requirements.
Therefore, we performed a dedicated analysis by grouping multiple adjacent array readout channels to study the expected performance with cells $2\!\times\!1.625$~mm$^2$, $3\!\times\!1.625$~mm$^2$ and $4\!\times\!1.625$~mm$^2$.
For this analysis, if a channel in a given group of readout channels had been fired, the center of the considered group was assumed for the X-Y coordinates of the hit position.
The considered analysis method allowed us to extract general predictions on the achievable single photon and ring angular resolution with the increasing size of readout pixels. The resulting angular distributions were fitted using the same procedure discussed for the actual pixel size. The results for the extrapolated single photon and ring angular resolution are shown in Fig.~\ref{fig:sigma_vs_pixel_size}.
For the single photon angular resolution $\sigma_c$, the main contributions are combined according to the relation

\begin{equation}
    \sigma^2_c = \sigma^2_{Trk} + \sigma^2_{Ep} + \sigma^2_{Chro} + \sigma^2_{Pix}
    \label{Eq:sigma_single_general}
\end{equation}
where $\sigma_{Trk}$ is the uncertainty related to charged particle tracking, $\sigma_{Ep}$ is the geometrical uncertainty related to the photon emission point, $\sigma_{Chro}$ is the uncertainty due to the chromatic dispersion of the radiator and $\sigma_{Pix}$ is the uncertainty introduced by the finite pixel size of the photodetector \cite{RICH_seguinot_ypsilantis}.
The expected dependence of $\sigma_c$ on the linear size of the readout pitch~$l$ of the ring arrays can be expressed writing Eq. \ref{Eq:sigma_single_general} in the form

\begin{equation}
    \sigma_c(l) = \sqrt{a^2 + b^2 l^2  }
    \label{Eq:sigma_single_vs_pixel_size}
\end{equation}
where $a$ includes the fixed contributions from $\sigma_{Trk}$, $\sigma_{Ep}$, $\sigma_{Chro}$ and array transverse size-related $\sigma_{Det}$ component, while $b$ weights the  $\sigma_{Pix}$ contribution from the hit localization uncertainty proportional to $l$.
From the fit of the measured points in Fig.\ref{fig:sigma_vs_pixel_size} with the function in Eq. \ref{Eq:sigma_single_vs_pixel_size}, it can be observed that the data are in agreement with the model.
The worsening of the single photon angular resolution is negligible, down to a few tenths of mrad, up to $l = 2$ mm, and becomes more and more relevant, beyond a few mrad, for larger $l$ values. The same scaling is observed for the ring angular resolution, requiring more detected ring photons to achieve a target resolution with the increasing $l$ while significantly reducing the number of required electronics channels. According to the results of these studies, with the baseline 2 mm SiPM pitch the target 1.5 mrad resolution should be still achieved for $\text{N}_{\text{ph}}>7$.

Finally, the scaling of the fiducial ring area where Cherenkov photons from different particles are clustered with the increasing $l$ was extrapolated. 
The ring area is a crucial design parameter to take into account for pattern recognition, especially in the high charged track multiplicity environment of central Pb-Pb collisions expected in ALICE~3, where hundreds of charged tracks reaching the radiator are expected per m$^2$. A limited ring area is mandatory to minimize the background from hits due to the SiPM DCR, electronics noise and photons from different tracks.
For this study, we evaluated the $\text{N}$ sigma fiducial ring area assuming full ring coverage with SiPMs as the area of the circular crown  with inner and outer radii $R_c - \text{N} \sigma_R$ and $R_c + \text{N} \sigma_R$, respectively. Here $R_c$ and $\sigma_R$ are the mean and sigma of the signal Gaussian fit component to radial distribution at single photon level.
The extrapolated values for $\text{N}$ = 1, 2 and 3 as a function of $l$ for pions at 10 GeV/$\it{c}$ momentum and protons at 8 GeV/$\it{c}$ momentum are shown in Fig. \ref{fig:area_ring_full_coverage}.
The three sigma ring area at saturation is approximately 25 cm$^2$ for $l=1$ mm and significantly increases with $l$.  
The observed increase of the fiducial ring area with $l$, together with dedicated Monte Carlo simulations aimed at studying the stability of pattern recognition in central Pb-Pb, highlights the importance of keeping a linear SiPM readout pitch not larger than 2 mm for our application, in agreement with the current ALICE~3 bRICH baseline specifications. 

\section{Conclusions}
A prototype for the future ALICE~3 bRICH detector, based on aerogel and SiPMs in a proximity-focusing layout, has been designed, assembled, and successfully tested at the CERN PS in October 2023. 
The results of our measurements based on commercial off-the-shelf electronic components validate the design bRICH specifications in terms of expansion gap, aerogel thickness and pixel size, as well as the aerogel refractive index to achieve the target angular resolution and separation power in the required momentum interval.  In particular, we measured a single photon angular resolution of 3.8 mrad at the Cherenkov angle saturation value of 242 mrad. The observed scaling of the resolution with the increasing number of detected ring photons is in excellent agreement with the 1.5 mrad bRICH design target.
We also studied the impact of uncorrelated and correlated background sources on the signal and proved the effectiveness of timing matching between tracks and hits in the background suppression. 
For the next measurements, the setup will be upgraded replacing the ring arrays with arrays equipped with the baseline 2x2~mm$^{2}$ SiPMs and providing larger ring acceptance for refined pattern recognition studies. The PETIROC 2A ASIC will be replaced by dedicated electronics to enhance the single-photon capabilities.
Measurements on irradiated arrays at the target -40~$^\circ$C operation temperature will be crucial to prove the stability of the reconstruction performance with the increasing uncorrelated DCR background expected for fluence of the order of 10$^{11}$ 1-MeV n$_{eq}$/cm$^2$.

\section*{Acknowledgments}
The authors would like to thank the INFN Bari staff for its contribution to the procurement and to the construction of the prototype. In particular, we thank D.~Dell'Olio, M.~Franco, N.~Lacalamita, F.~Maiorano, M.~Mongelli, M.~Papagni, C.~Pastore and R.~Triggiani for their technical support.
The authors also acknowledge the ALICE Collaboration and the T10 CERN team for providing the facilities and the support to the organization of the beam test throughout all the beam test duration.
The research leading to these results received funding from the European Union's Horizon Europe research and innovation programme under grant agreement No. 101057511.

\bibliographystyle{unsrt}
\bibliography{RichPaper1}

\end{document}